\documentclass{ws-ijmpa}
\usepackage[super,compress]{cite}
\usepackage{bm}
\usepackage{hyperref}
\usepackage{listings}
\usepackage{color}
\usepackage{pstricks,psfrag,pst-node,pst-text,pst-3d,pst-grad}

\newcommand{\dif}{{\rm d}}
\newcommand{\del}{\partial}
\newcommand{\lie}{\pounds\!}
\newcommand{\tdif}{\tilde{D}}
\newcommand{\tgam}{\tilde{\gamma}}
\newcommand{\tGam}{\tilde{\Gamma}}
\newcommand{\tlap}{\tilde{\bigtriangleup}}
\newcommand{\tA}{\tilde{A}}
\newcommand{\tR}{\tilde{\mathcal{R}}}
\newcommand{\sR}{\mathcal{R}}
\newcommand{\ts}{\tilde{s}}
\newcommand{\tr}{\tilde{r}}
\newcommand{\br}{\bar{r}}
\newcommand{\bs}{\bar{s}}
\newcommand{\mJ}{\mathcal{J}}
\newcommand{\mGS}{\mathcal{GS}}
\newcommand{\mS}{\mathcal{S}}
\newcommand{\mR}{\mathcal{R}}
\newcommand{\mP}{\mathcal{P}}

\begin{document}


%
\catchline{}{}{}{}{}
%

\title{Initial Conditions for Numerical Relativity\\
$\sim$ Introduction to numerical methods for solving elliptic PDEs $\sim$
}

\author{Hirotada Okawa
\footnote{hirotada.okawa@ist.utl.pt}
}

\address{CENTRA, Departamento de F\'{\i}sica, Instituto Superior T\'ecnico,\\
Universidade T\'ecnica de Lisboa - UTL,
Av.~Rovisco Pais 1, 1049 Lisboa, Portugal.\\
hirotada.okawa@ist.utl.pt}

\maketitle


\begin{abstract}
 Numerical relativity became a powerful tool to investigate
 the dynamics of binary problems with black holes or neutron stars as well as the very structure
 of General Relativity. Although public numerical relativity codes are available to evolve such systems, a proper understanding
 of the methods involved is quite important. Here we focus on the numerical solution of elliptic partial differential equations.
 Such equations arise when preparing initial data for numerical relativity, but also for monitoring the evolution of black holes.
 Because such elliptic equations play an important role in many branches of physics, we give an overview of the topic,
 and show how to numerically solve them with simple examples
 and sample codes written in C++ and Fortran90 for beginners in numerical relativity or other fields requiring numerical expertise.
\end{abstract}

\ccode{PACS numbers:}

\section{Introduction}
Numerical relativity is now a mature science, the purpose of which is to investigate non-linear dynamical
spacetimes. A traditional example of the application and importance of numerical relativity concerns the modelling of gravitational-wave emission and
consequent detection. In order to detect gravitational waves(GWs) from black hole-neutron star (BH-NS), BH-BH or NS-NS binaries, one needs to
accurately understand the waveforms from these sources in advance, because their signals are quite faint
for our detectors~\cite{Sathyaprakash2009}.
To understand why the problem is so difficult, consider Newtonian gravity, as applied to systems like our very own Earth-Moon.
In Newton's theory, binary systems can move on stable, circular or quasi-circular orbits.
However, in binary systems heavy enough or moving sufficiently fast, 
the effects of General Relativity become important, and the notion of stable orbits is no longer valid:
GWs take energy and angular momentum away from the system and energy conservation implies that the binary orbit shrinks until finally the objects merge and presumably form a final single object. Accordingly, the evolution of binary stars can be divided into an
inspiral, merger and a ring-down phase\cite{Pretorius2007}.

In the inspiral phase, GW emission is sufficiently under control by using slow-motion, Post-Newtonian expansions because the stars are distant from each other and their gravitational forces can be described in a
perturbation scheme~\cite{Blanchet2002,Buonanno1998}. The ringdown phase describes the vibrations of the final object.
Because of the uniqueness theorems~\cite{Carter1997}, GWs can be computed by
BH perturbation methods~\cite{Kokkotas1999,Berti:2009kk}. Advanced BH perturbation methods are reviewed in Ref.~\cite{Pani:2013pma}.
Numerical Relativity enables us to obtain the GW form in all phases~\cite{Boyle2007,Shibata2011,Faber2012}.

Furthermore, we note that techniques of numerical relativity are also available in a variety of contexts.
For example, but by no means the only one, it became popular to investigate the nature of higher
dimensional spacetimes~\cite{Cardoso2012}, most specially in the framework of large extra dimensions~\cite{ArkaniHamed1998r,Antoniadis1998i,Randall1999v,Randall1999e}.
It was pointed out that a micro BH can be produced from high energy particle collisions at the Large Hadron Collider(LHC) and beyond~\cite{Dimopoulos2001,Giddings2001}, and while some works use shock wave collisions~\cite{Sampaio:2013faa,Penrose1974,Eardley2002}
the full-blown numerical solution is clearly desirable to investigate the nature of gravity
with high energy collisions in four dimensional spacetime\cite{Shibata2008,Sperhake2008,Sperhake2009,East2012}.
If we consider spacetime dimensions higher than four in our simulations, larger computational resources will be required.
However, it can be reduced to four dimensional problem with small
changes assuming the symmetry of spacetimes~\cite{HelviLec,Yoshino2009,Witek2010x,Witek2010a,Zilhao2011,Okawa2011}.
As a result, to investigate the nature of higher dimensional spacetimes
is in the scope of numerical relativity~\cite{Maliborski:2013via,Shibata2010,Lehner2011}.

Fortunately, open source codes to evolve dynamical systems
with numerical relativity are available~\cite{Zilhao:2013hia,cactusweb,lorene,EinsteinToolkitweb}.
All that one needs to do is to prepare the initial data describing the physics of the problem one is interested in.
Here, we explain precisely how this is achieved, to prepare initial data for
numerical relativity in this paper.
Briefly, it amounts to solving an elliptic partial differential equation (PDE)
and we explain how to solve the elliptic PDE from the numerical point of
view of beginners in numerical studies.



\section{ADM formalism}
In numerical relativity, we regard our spacetimes as the evolution of spaces.
We begin by showing how to decompose the spacetime into
timelike and spacelike components in the ADM formalism\cite{Arnowitt1962,Misner1974,Wald1984}.
Then, we derive evolution equations from Einstein's
equations along the lines of York's review~\cite{York1979}.

\subsection{Decomposition}
First, we introduce a family of three-dimensional spacelike hypersurfaces $\Sigma$
in four-dimensional manifold $V$.
Hypersurfaces $\Sigma$ are expressed as the level surfaces of a scalar function~$f$
and are not supposed to intersect one another.
We can define a one-form $\Omega_{\mu}=\nabla_{\mu}f$ which is normal to
the hypersurface.

Let $g_{\mu\nu}$ be a metric tensor in four-dimensional manifold $V$.
The norm of one-form $\Omega_{\mu}$ can be written by a positive
function $\alpha$ as
\begin{eqnarray}
 g^{\mu\nu}\Omega_{\mu}\Omega_{\nu} = -\frac{1}{\alpha^2},
\end{eqnarray}
where $\alpha$ is called lapse function.
We define a normalized one-form by
\begin{eqnarray}
 \omega_{\mu} =\alpha\Omega_{\mu},\ \ \ g^{\mu\nu}\omega_{\mu}\omega_{\nu}=-1.
\end{eqnarray}
The orthogonal vector to a hypersurface $\Sigma$ is written by
\begin{eqnarray}
 n^{\mu}=-g^{\mu\nu}\omega_{\nu},
\end{eqnarray}
whose minus sign is defined to direct at the future
and we note that this timelike vector satisfies $n^{\mu}n_{\mu}=-1$ by definition.

\subsubsection{Induced metric}
The induced metric $\gamma_{\mu\nu}$ on $\Sigma$
and the projection tensor $\perp_{\nu}^{\mu}$ from $V$ to $\Sigma$
are given by the four-dimensional metric $g_{\mu\nu}$,
\begin{eqnarray}
 \gamma_{\mu\nu} &=& g_{\mu\nu} +n_{\mu}n_{\nu},\\
 \perp^{\mu}_{\nu} &=& \delta^{\mu}_{\nu} +n^{\mu}n_{\nu},
\end{eqnarray}
where one can show $n^{\mu}\gamma_{\mu\nu}=0$,
which yields that timelike components of $\gamma_{\mu\nu}$ vanish
and only spacelike components~$\gamma_{ij}$ exist.
The induced covariant derivative $D_{i}$
on $\Sigma$ is also defined in terms of the four-dimensional covariant
derivative $\nabla_{\mu}$.
\begin{eqnarray}
 D_{i} \psi &=& \perp^{\rho}_{i}\nabla_{\rho}\psi,\\
 D_{j} W^{i} &=& \perp^{\rho}_{j}\perp^{i}_{\lambda}\nabla_{\rho}W^{\lambda},
\end{eqnarray}
where $\psi$ and $W^{\lambda}$ denote arbitrary scalar and vector on $\Sigma$.
By a straightforward calculation, one can show that the induced
covariant derivative satisfies $D_{i}\gamma_{jk}=0$.

\subsubsection{Curvature}
Riemann tensor on $\Sigma$ is defined using an arbitrary vector $W^i$ by
\begin{eqnarray}
 D_{[i}D_{j]}W_k &=& \frac{1}{2}\sR_{ijk}^{\quad\,\ell}W_{\ell},\\
 \sR_{ijk\ell}n^{\ell} &=& 0,
\end{eqnarray}
where $[\ ]$ denotes the antisymmetric operator for indices
and $\sR_{ijk\ell}$ denotes Riemann tensor on $\Sigma$.
Ricci tensor is determined by the contraction
of the induced metric and Riemann tensor on $\Sigma$.
Ricci scalar is also determined by the contraction
of the induced metric and Ricci tensor.

We define the extrinsic curvature on $\Sigma$, which
describes how the hypersurface is embedded in the manifold $V$.
The extrinsic curvature is defined by
\begin{eqnarray}
 K_{\mu\nu} = -\perp^{\rho}_{\mu}\perp^{\lambda}_{\nu}
  \nabla_{(\rho}n_{\lambda )},
\end{eqnarray}
where $(\,)$ denotes the symmetric operator for indeces.
One can also show that the extrinsic curvature is spacelike
by multiplying the normal vector $n^{\mu}$ 
in the same manner as $\gamma_{\mu\nu}$.
In addition,
by the definition of the projection tensor, we obtain the following relation between the covariant derivative of the
normal vector and their projection,
\begin{eqnarray}
 \nabla_{\mu}n_{\nu} &=& \left(\perp_{\mu}^{\rho} -n_{\mu}n^{\rho}\right)
  \left(\perp_{\nu}^{\lambda} -n^{\lambda}n_{\nu}\right)
  \nabla_{\rho}n_{\lambda}\nonumber\\
 &=& \perp_{\mu}^{\rho}\perp_{\nu}^{\lambda} \nabla_{\rho}n_{\lambda} -\perp_{\nu}^{\lambda}n_{\mu}n^{\rho}\nabla_{\rho}n_{\lambda}
  -\perp_{\mu}^{\rho}n_{\nu}n^{\lambda}\nabla_{\rho}n_{\lambda}
  +n_{\mu}n_{\nu}n^{\rho}n^{\lambda}\nabla_{\rho}n_{\lambda}\nonumber\\
 &=& \perp_{\mu}^{\rho}\perp_{\nu}^{\lambda} \nabla_{\rho}n_{\lambda} -n_{\mu}n^{\rho}\nabla_{\rho}n_{\nu},
\end{eqnarray}
where the relation $n^{\lambda}n_{\lambda}=-1$ is used in the last equation.
Then, the extrinsic curvature can be rewritten by
\begin{eqnarray}
 K_{\mu\nu} &=& -\frac{1}{2}
  \left\{\nabla_{\mu}n_{\nu}
   +\nabla_{\nu}n_{\mu}+n_{\mu}n^{\rho}\nabla_{\rho}n_{\nu}
   +n_{\nu}n^{\rho}\nabla_{\rho}n_{\mu}\right\}\nonumber\\
 &=& -\frac{1}{2}
  \left\{
   \gamma_{\nu\rho}\nabla_{\mu}n^{\rho} +\gamma_{\mu\rho}\nabla_{\nu}n^{\rho}
   +n^{\rho}\nabla_{\rho}\gamma_{\mu\nu}
  \right\}\nonumber\\
 &=& -\frac{1}{2}\lie_{n}\gamma_{\mu\nu},
\end{eqnarray}
where $\lie_{n}\gamma_{\mu\nu}$ denotes the Lie derivative of the tensor $\gamma_{\mu\nu}$ along the vector $n^{\mu}$.
The geometrical nature of the three-dimensional hypersurfaces
can be determined by the induced metric and extrinsic curvature on $\Sigma$.
$K_{ij}$ and $\gamma_{ij}$ must satisfy the following geometrical relations
to embed $\Sigma$ in $V$.

\subsubsection{Geometrical relations}
We derive geometrical relations by the projection of
the four-dimensional Riemann tensor to the hypersurface $\Sigma$.
First, in order to obtain the relation between
the four-dimensional Riemann tensor~ $R_{\mu\nu\rho\lambda}$
and the three-dimensional Riemann tensor~ $\sR_{ijk\ell}$,
we rewrite the definition of $\sR_{ijk\ell}$
 with $R_{\mu\nu\rho\lambda}$ and the extrinsic curvature,
\begin{eqnarray}
 \perp^{\mu}_{i}\perp^{\nu}_{j}\perp^{\rho}_{k}\perp^{\lambda}_{\ell} R_{\mu\nu\rho\lambda}
  = \sR_{ijk\ell} +K_{ik}K_{j\ell} -K_{jk}K_{i\ell}.\label{eq:gauss_rel}
\end{eqnarray}
Eq.~(\ref{eq:gauss_rel}) is called Gauss' equation.
Secondly, we project the four-dimensional Riemann tensor
contracted by an orthogonal normal vector $n^{\lambda}$.
\begin{eqnarray}
 \perp^{\mu}_{i}\perp^{\nu}_{j}\perp^{\rho}_{k}R_{\mu\nu\rho\lambda}n^{\lambda}
  = D_{j}K_{ik} -D_{i}K_{jk}.\label{eq:codazzi_rel}
\end{eqnarray}
Eq.~(\ref{eq:codazzi_rel}) is called Codazzi's equation.
Finally, we consider a Lie derivative of the extrinsic curvature
to the time direction.
We define a timelike vector $t^{\mu}$
with a lapse function and a shift vector $\beta^{\mu}$
which satisfies $\Omega_{\mu}\beta^{\mu}$ as
\begin{eqnarray}
 t^{\mu} = \alpha n^{\mu} +\beta^{\mu}.
\end{eqnarray}
Then, with the Lie derivative along $t^{\mu}$ and $\beta^{\mu}$,
we rewrite the four dimensional Riemann tensor
contracted by two orthogonal normal vectors as
\begin{eqnarray}
 \perp^{\mu}_{i}\perp^{\nu}_{j}R_{\mu\rho\nu\lambda}n^{\rho}n^{\lambda} 
  = \frac{1}{\alpha}\left[\lie_{t}-\lie_{\beta}\right]K_{ij}
  -K_{i\ell}K^{\ell}_{j} -\frac{1}{\alpha}D_{i}D_{j}\alpha,\label{eq:ricci_rel}
\end{eqnarray}
which Eq.~(\ref{eq:ricci_rel}) is called Ricci's equation.

\subsection{Decomposition of Einstein's equations}
\label{sec:decomposition_einstein}
Let us now use the geometric relations to decompose Einstein's equations.
Let us for convenience define the Einstein tensor $G_{\mu\nu}$,
\begin{eqnarray}
 G_{\mu\nu}\equiv R_{\mu\nu}-\frac{1}{2}g_{\mu\nu}R
  = \frac{8\pi G}{c^4}T_{\mu\nu},\label{eq:einstein0}
\end{eqnarray}
where $G$ denotes the gravitational constant and $c$ denotes the speed
of light and hereafter we set $G=c=1$ for simplicity.
We start by decomposing the energy momentum tensor as
\begin{eqnarray}
 T_{\mu\nu} = S_{\mu\nu} +2j_{(\mu}n_{\nu)}
  +\rho n_{\mu}n_{\nu},
\end{eqnarray}
where $\rho \equiv T_{\mu\nu}n^{\mu}n^{\nu}$,
$j_{\mu}\equiv -\perp^{\rho}_{\mu}T_{\rho\lambda}n^{\lambda}$
and $S_{\mu\nu}\equiv\perp^{\rho}_{\mu}\perp^{\lambda}_{\nu}T_{\rho\lambda}$.

We multiply Gauss' equation~(\ref{eq:gauss_rel}) by
an induced metric~ $\gamma^{ik}$ and obtain
\begin{eqnarray}
 \perp^{\nu}_{j}\perp^{\lambda}_{\ell}
  \left[R_{\nu\lambda}-R_{\mu\nu\rho\lambda}n^{\mu}n^{\rho}\right]
  = \sR_{j\ell} +KK_{j\ell} -K^{\, i}_{j}K_{i\ell}.\label{eq:gauss_rel_gik}
\end{eqnarray}
In addition, Eq.~(\ref{eq:gauss_rel_gik}) contracted by $\gamma^{j\ell}$
gives twice as much as the Einstein's tensor
contracted by two orthogonal normal vectors~$n^{\mu}$~and~$n^{\nu}$.
Then, we obtain
\begin{eqnarray}
 \sR +K^2-K_{ij}K^{ij} = 16\pi\rho.\label{eq:ham_const_0}
\end{eqnarray}
Similarly, Codazzi's equation~(\ref{eq:codazzi_rel}) contracted by $\gamma^{jk}$ results in
\begin{eqnarray}
 D_{j}K_{i}^{\, j} -D_{i}K = 8\pi j_{i}.\label{eq:mom_const_0}
\end{eqnarray}
Note that Eq.~(\ref{eq:ham_const_0}) and Eq.~(\ref{eq:mom_const_0})
are composed of only spacelike variables and should be satisfied on
each hypersurface $\Sigma$ because they do not depend on time.
Therefore, Eq.~(\ref{eq:ham_const_0}) and Eq.~(\ref{eq:mom_const_0})
are called the Hamiltonian and momentum constraints, respectively.

Finally, let us rewrite Ricci's equation~(\ref{eq:ricci_rel}).
Einstein's equations~(\ref{eq:einstein0}) can also be expressed with
the trace of the energy momentum tensor $T\equiv g^{\mu\nu}T_{\mu\nu}$ as
\begin{eqnarray}
 R_{\mu\nu} = 8\pi \left[T_{\mu\nu}-\frac{1}{2}g_{\mu\nu}T\right].\label{eq:einstein1}
\end{eqnarray}
The projection of Einstein's equations~(\ref{eq:einstein1})
yields
\begin{eqnarray}
 \perp^{\mu}_{i}\perp^{\nu}_{j}R_{\mu\nu}
  = 8\pi\left[S_{ij}-\frac{1}{2}\gamma_{ij}\left(S-\rho\right)\right],\label{eq:einstein2}
\end{eqnarray}
where $S=\gamma^{ij}S_{ij}$.
Ricci's equation~(\ref{eq:ricci_rel})
is rewritten with Eq.~(\ref{eq:gauss_rel_gik})~and Eq.~(\ref{eq:einstein2}) as
\begin{eqnarray}
 \lie_{t}K_{ij} &=& \lie_{\beta}K_{ij}
  +\alpha\left(\sR_{ij}-2K_{i\ell}K^{\, \ell}_{j}+K_{ij}K\right)
 -D_{j}D_{i}\alpha\nonumber\\
 & &  
  -8\pi\alpha\left[S_{ij}+\frac{1}{2}\left(\rho -S\right)\gamma_{ij}\right].
\end{eqnarray}

\subsection{Propagation of Constraints}
\label{sec:propagation_const}
In ADM formalism, Einstein's equations are regarded as evolution equations in time with
the geometrical constraints on each hypersurface.
In general, it is numerically expensive to guarantee the constraints on
each step because we must solve elliptic PDEs as described in Sec.~\ref{sec:solve_constr}.
However, in principle, one does not have to solve the constraint equations
if the initial data satisfy the constraints~\cite{Frittelli1997,Yoneda2001,Alcubierre2008}.
This works as follows.
We first define the following quantities,
\begin{eqnarray}
 C&\equiv& \left(G_{\mu\nu}
  -8\pi T_{\mu\nu}\right)n^{\mu}n^{\nu},\label{eq:prop_hamiltonian}\\
 C_{\mu}&\equiv& -\perp_{\mu}^{\rho} \left(G_{\rho\nu}
  -8\pi T_{\rho\nu}\right)n^{\nu},\label{eq:prop_momentum}\\
 C_{\mu\nu}&\equiv& \perp_{\mu}^{\rho}\perp_{\nu}^{\lambda}
  \left(G_{\rho\lambda}
  -8\pi T_{\rho\lambda}\right),\label{eq:prop_evolution}
\end{eqnarray}
where $C=0$ corresponds to the Hamiltonian contstraint,
$C_{\mu}=0$ correspond to the momentum constraints and
$C_{\mu\nu}=0$ denote the evolution equations in ADM formalism.
Einstein's equations can be decomposed in terms of
$C,C_{\mu}$ and $C_{\mu\nu}$ as
\begin{eqnarray}
 G_{\mu\nu}-8\pi T_{\mu\nu}&=&
  \left(\perp_{\mu}^{\rho}-n_{\mu}n^{\rho}\right)
  \left(\perp_{\nu}^{\lambda}-n_{\nu}n^{\lambda}\right)
  \left(G_{\rho\lambda} -8\pi T_{\rho\lambda}\right)\nonumber\\
&=& C_{\mu\nu}+n_{\nu}C_{\mu}+n_{\mu}C_{\nu}+n_{\mu}n_{\nu}C.
\label{eq:prop_einstein}
\end{eqnarray}
Thanks to the Bianchi identity which is a mathematical relation for the Riemann tensor,
the covariant derivative of Einstein's tensor vanishes.
Besides, the covariant derivative of the energy momentum tensor
which appears in the right-hand-side of Einstein's equations denotes the
energy conservation law,
\begin{eqnarray}
 \nabla^{\nu}\left(G_{\mu\nu}-8\pi T_{\mu\nu}\right)&=& 0.
\end{eqnarray}
Let us project the covariant derivative of Einstein's equations
to $n^{\mu}$ direction and to the spatial direction with $\perp^{\rho}_{\mu}$.
\begin{eqnarray}
 n^{\mu}\nabla^{\nu}\left(G_{\mu\nu}-8\pi T_{\mu\nu}\right)&=&
  -C_{\mu\nu}D^{\nu}n^{\mu} -2C_{\mu}n^{\nu}\nabla_{\nu}n^{\mu} -D^{\mu}C_{\mu}
  -C \nabla_{\mu}n^{\mu}-n^{\mu}\nabla_{\mu}C,
\nonumber\\
\label{eq:deinstein_normal}\\
 \perp^{\rho}_{\mu}\nabla^{\nu}\left(G_{\rho\nu}-8\pi T_{\rho\nu}\right)&=&
  D^{\nu}C_{\nu\mu}+C_{\mu\rho}n^{\nu}\nabla_{\nu}n^{\rho}
  +2n^{\nu}\nabla_{\nu}C_{\mu}
  -C_{\rho}n_{\mu}n^{\nu}\nabla_{\nu}n^{\rho},
\label{eq:deinstein_spatial}
\end{eqnarray}
where $D_\mu$ denotes the covariant derivative with respect to
the induced metric, noting that $C_{\mu}$ and $C_{\mu\nu}$ are spatial.
Thus, we show the propagation of constraints along the timelike vector as
\begin{eqnarray}
 n^{\mu}\nabla_{\mu}C&=&
  -C_{\mu\nu}D^{\nu}n^{\mu} 
  -2C_{\mu}n^{\nu}\nabla_{\nu}n^{\mu}
  -D^{\mu}C_{\mu} -C \nabla^{\mu}n_{\mu},
  \label{eq:propagation_hamiltonian}\\
 n^{\nu}\nabla_{\nu}C_{\mu} &=&
  -\frac{1}{2}D^{\nu}C_{\mu\nu} -\frac{1}{2}C_{\mu\rho}n^{\nu}\nabla_{\nu}n^{\rho}
  +\frac{1}{2}C_{\rho}n_{\mu}n^{\nu}\nabla_{\nu}n^{\rho}. 
\label{eq:propagation_momentum}
\end{eqnarray}
Therefore, we can keep the Hamiltonian and momentum constraints
satisfied during evolution,
as long as we evolve the initial data satisfying the constraints 
$C=0$ and $C_{\mu}=0$ on $\Sigma$ by the evolution equation $C_{\mu\nu}=0$.
It should be noted that the ADM evolution equations are numerically unstable and not suitable for numerical evolutions;
instead one uses hyperbolic evolution equations for time integration,
for example, BSSN and Z4 evolution equations~\cite{DavidLec,HelviLec,Shibata1995,Baumgarte1998,Bona2003fj}.

\section{Initial Condition for Numerical Relativity}
As emphasized in Sec.~\ref{sec:decomposition_einstein},
initial data cannot be freely specified, as it needs to satisfy the Hamiltonian
and momentum constraints on a hypersurface $\Sigma$. 
In general, the problem of constructing initial data
is called ``Initial Value Problem''.
The standard method for solving an initial value problem
is reviewed in Ref.\cite{Cook2000,Gourgoulhon2007}.
In this section, we derive the equations for the initial value problem
and then introduce some examples as initial data for numerical relativity.

\subsection{Initial Value Problem}
\label{sec:solve_constr}
There are twelve variables($\gamma_{ij}, K_{ij}$) as the metric part 
to be determined and four constraint equations in ADM formalism.
One can obtain four variables by solving constraints after assuming
eight variables by physical and numerical reasons.

\subsubsection{York-Lichnerowicz conformal decomposition}
To begin with, we introduce the conformal factor $\psi$ as
\begin{eqnarray}
 \gamma_{ij} &=& \psi^4\tgam_{ij},
\end{eqnarray}
where we define $\det\tgam_{ij}\equiv 1$ and we have one degree of freedom
in $\psi$ and five degrees of freedom in $\tgam_{ij}$.
By the conformal transformation,
the following relations between variables with respect to $\gamma_{ij}$
and $\tgam_{ij}$ are immediately given by
\begin{eqnarray}
 \Gamma^i_{jk} &=& \tGam^i_{jk} +\frac{2}{\psi}
  \left[\delta^i_j\tdif_k\psi +\delta^i_k\tdif_j\psi
 -\tgam^{il}\tgam_{jk}\tdif_l\psi\right],\\
 \sR &=& \tR\psi^{-4} -8\psi^{-5}\tlap\psi,
\end{eqnarray}
where $\tdif_i,\tGam^i_{jk},\tR$ and $\tlap$ are respectively 
covariant derivative, Ricci scalar, Christoffel
symbol and Laplacian operator defined by $\tlap\psi=\tgam^{ij}\tdif_i\tdif_j\psi$
with respect to $\tgam_{ij}$.

\subsubsection{Transverse-Traceless decomposition}
As for the extrinsic curvature, we start by decomposing it into
a trace and a trace-free part,
\begin{eqnarray}
 K_{ij} &=& A_{ij} +\frac{1}{3}\gamma_{ij}K,
\end{eqnarray}
where $\gamma^{ij}A_{ij}=0$ and $K=\gamma^{ij}K_{ij}$ and we have one
degree of freedom in $K$ and five degrees of freedom in $A_{ij}$.
Then, we also define the conformal transformation for $A_{ij}$ as
\begin{eqnarray}
 A_{ij} &=& \psi^{-2}\tA_{ij}.
\end{eqnarray}
According to the definition of derivatives with respect to $\gamma_{ij}$
and $\tgam_{ij}$, we obtain the following relation,
\begin{eqnarray}
 D_jA^{ij} &=& \psi^{-10}\tdif_i\tA^{ij}.
\end{eqnarray}
In addition, we decompose the conformal traceless tensor $\tA_{ij}$
into a divergenceless part and a ``derivative of a vector'' $W_j$ part.
Hereafter we assume the divergenceless part vanishes for simplicity.
The conformal traceless extrinsic curvature is described by
\begin{eqnarray}
 \tA_{ij} &=& \tdif_iW_j +\tdif_jW_i -\frac{2}{3}\tgam_{ij}\tdif_kW^k.\label{eq:def_taij}
\end{eqnarray}
The covariant derivative of $\tA_{ij}$ is written by
\begin{eqnarray}
 \tdif_i\tA^i_j &=& \tlap W_j 
  +\tdif_i\tdif_jW^i
  -\frac{2}{3}\tdif_j\tdif_kW^k\nonumber\\
  &=& \tlap W_j 
  +\frac{1}{3}\tdif_j\tdif_kW^k
  +\tR_{ij}W^i,
\end{eqnarray}
where we used the definition of Riemann tensor.

\subsubsection{Constraints as initial value problem}
With the above conformal transformation, the Hamiltonian and momentum constraints are rewritten as
\begin{eqnarray}
 \tlap\psi -\frac{1}{8}\tR\psi+\frac{1}{8}\tA_{ij}\tA^{ij}\psi^{-7}
  -\frac{1}{12}K^2\psi^5 &=& 16\pi\psi^5\rho,\label{eq:ham_const_ivp}\\
 \tlap W_i +\frac{1}{3}\tdif_i\tdif_k W^k +\tR_{ij}W^j
  -\frac{2}{3}\psi^6\tdif_iK&=& 8\pi\psi^6 j_i.\label{eq:mom_const_ivp}
\end{eqnarray}

\subsection{Schwarzschild Black Hole}
Let us consider an exact solution of Einstein's equations as initial data
for numerical relativity.
The Schwarzschild BH is the simplest BH solution in static and spherically
symmetric spacetimes~\cite{Schwarzschild1916}.
The line element of the Schwarzschild BH in spherical
coordinates~$(\br, \theta, \phi)$ is given by
\begin{eqnarray}
 \dif s^2 = -f_0\dif t^2 +f^{-1}_0\dif \br^2
  +\br^2\left(\dif\theta^2 +\sin^2\theta\dif\phi^2\right),
\end{eqnarray}
where we define $f_0$ with BH mass $M$,
\begin{eqnarray}
 f_0(\br)=1-\frac{2M}{\br}.
\end{eqnarray}
Let us define the coordinate transformation by
\begin{eqnarray}
 \br^2 &\equiv& \psi_0^4r^2,\\
 \frac{\dif \br^2}{1-\frac{2M}{\br}} &\equiv& \psi^4_0\dif r^2,
\end{eqnarray}
where $r$ denotes the isotropic radial coordinate and we introduce
a scalar function~ $\psi_0$.
Then, we solve $\br$ under the coordinate transformation and
obtain $\psi_0$ and the relation between $\br$ and $r$ as
\begin{eqnarray}
 \psi_0 &=& 1+\frac{M}{2r},\\ 
 \frac{\dif \br}{\dif r} &=& \left(1+\frac{M}{2r}\right)\left(1-\frac{M}{2r}\right).
\end{eqnarray}
After straightforward calculations, the line element is rewritten by
\begin{eqnarray}
 \dif s^2 &=& -\left(\frac{1-\frac{M}{2r}}{1+\frac{M}{2r}}\right)^2\dif t^2
  +\left(1+\frac{M}{2r}\right)^4
  \left[\dif r^2 +r^2\dif\theta^2 +r^2\sin^2\theta\dif\phi^2\right]\nonumber\\
  &=& -\alpha_0^2\dif t^2 +\psi_0^4 \eta_{ij}\dif x^i\dif x^j,\label{eq:sbh_iso}
\end{eqnarray}
where $\eta_{ij}$ denotes the flat metric and we define $\alpha_0$.
In the isotropic coordinates, all spatial metric components remain regular,
in contrast to the ones in the standard Schwarzschild coordinates.
The range $[2M<\br <\infty]$ in the spherical coordinates
corresponds to $[\frac{M}{2}<r<\infty]$
in the isotropic coordinates.
In addition, when we change to a new coordinate $\tr\equiv \left(M/2\right)^2/r$,
we obtain the same expression as Eq.~(\ref{eq:sbh_iso})
with $\tr$ instead of $r$.
It yields that
the range $[\frac{M}{2}<\tr <\infty]$ corresponds to $[0<r<\frac{M}{2}]$.
The solution is inversion symmetric at $r=M/2$ and corresponds to the Einstein-Rosen bridge\cite{Einstein1935}.

Obviously, the extrinsic curvature $K_{ij}$ of the Schwarzschild BH
vanishes because the spacetime is static,
and therefore the momentum constraints are trivially satisfied.
Besides, the Hamiltonian constraint is also satisfied as
\begin{eqnarray}
 \bigtriangleup\psi_0&=&0,\label{eq:ham_const_sbh}
\end{eqnarray}
because the Schwarzschild BH is an exact solution of Einstein's equations.

\subsection{Puncture Initial Data}
One can analytically solve the momentum constraints with the following conditions,
\begin{eqnarray}
 \begin{array}{cccll}
  K &=&0&,&{\rm maximal\, condition},\\
  \tgam_{ij}&=& \eta_{ij}&,& \small {\rm conformal\, flatness},\\
  \psi\mid_{\infty} &=& 1&,& {\rm asymptotically\, flatness}\,.
 \end{array}\label{eq:punc_condition}
\end{eqnarray}
The derivative operator becomes quite simple assuming conformal flatness.
We also note that Eq.~(\ref{eq:ham_const_ivp}) and Eq.~(\ref{eq:mom_const_ivp}) are decoupled
with $K=const.$ condition.

\subsubsection{Single Black Hole}
Next, let us consider a BH with non-zero momentum($P^i\neq 0$), for
which the momentum constraints become non-trivial.
However, a solution to conditions~(\ref{eq:punc_condition}) can still be found.
In this case, the momentum constraints are given by
\begin{eqnarray}
 \bigtriangleup W_i +\frac{1}{3}\partial_i\partial_k W^k &=& 0.\label{eq:mom_const_ivp_punc}
\end{eqnarray}
We have a simple solution to satisfy Eq.~(\ref{eq:mom_const_ivp_punc}) as
\begin{eqnarray}
 W_i &=& -\frac{1}{4r}\left[7P_i+n_in_jP^j\right]
  +\frac{1}{r^2}\epsilon_{ijk}n^jS^k,\label{eq:sol_wi}
\end{eqnarray}
where $P^i$ and $S^i$ are constant vectors corresponding to the momentum and
spin of BH and $n^i\equiv x^i/r$ denotes the normal vector.
Then, we obtain the Bowen-York extrinsic curvature\cite{Bowen1980} by substituting Eq.~(\ref{eq:sol_wi})
into Eq.~(\ref{eq:def_taij}),
\begin{eqnarray}
 \tA^{(BY)}_{ij} &=& \frac{3}{2r^2}\left[P_in_j +P_jn_i -\left(\eta_{ij}-n_in_j\right)P^kn_k\right]
  +\frac{3}{r^3}\left[\epsilon_{kil}S^ln^kn_j +\epsilon_{kjl}S^ln^kn_i\right].\nonumber\label{eq:ext_by}\\
\end{eqnarray}

On the other hand, to satisfy the Hamiltonian
constraint~(\ref{eq:ham_const_ivp}) we must in general solve an elliptic PDE,
even if simple-looking,
\begin{eqnarray}
 \bigtriangleup\psi &=& -\frac{1}{8}\tA^{(BY)}_{ij}\tA_{(BY)}^{ij}\psi^{-7}.\label{eq:ham_const_punc0}
\end{eqnarray}
Let us define the function $u$ as a correction term relative to the Schwarzschild BH,
\begin{eqnarray}
 \psi &=& 1 +\frac{M}{2r} +u.
\end{eqnarray}
We can regularize the Hamiltonian constraint~(\ref{eq:ham_const_punc0})
when $ru$ is regular at the origin.
$\tA_{ij}$ is at most proportional to $r^{-3}$ at and $\psi$ is proportional to $r^{-1}$, so that
the divergent behavior of $\tA_{ij}\tA^{ij}$ is compensated by the $\psi^{-7}$ term at the origin.

Therefore, the Hamiltonian constraint yields
\begin{eqnarray} 
 \bigtriangleup u &=& -\frac{1}{8}\tA^{(BY)}_{ij}\tA_{(BY)}^{ij}\psi^{-7}.\label{eq:ham_const_punc}
\end{eqnarray}

\subsubsection{Multi Black Holes}
We can easily prepare the initial data which contains many BHs without any momenta
under the condition~(\ref{eq:punc_condition})
because the Hamiltonian constraint is the same as Eq.~(\ref{eq:ham_const_sbh})
and we know that the following conformal factor satisfies the Laplace equation.
\begin{eqnarray}
 \psi_{M} &=& 1+\sum_{n=1}^{N}\frac{M_n}{2\mid\bm{x}-\bm{x}_n\mid},
\end{eqnarray}
where $M_n$ and $\bm{x}_n$ denote the mass and position of n-th BH, respectively.
The initial data defined with $\psi=\psi_{M}$ and $\tA_{ij}=0$ is called
Brill-Lindquist initial data\cite{Brill1963}.

As for BHs with non-zero momenta,
we also use the Bowen-York extrinsic curvature and the same method for the
Hamiltonian constraint as $\psi\equiv\psi_M+u$,
\begin{eqnarray} 
 \bigtriangleup u &=& -\frac{1}{8}\psi^{-7}\sum_{n=1}^{N}\tA^{(BY,n)}_{ij}\tA_{(BY,n)}^{ij}.\label{eq:ham_const_puncm}
\end{eqnarray}
In principle, it is possible to construct initial data for multi BHs with any
momenta and spins by solving an elliptic PDE\cite{Brandt1997x}.

\subsection{Kerr Black Hole}
\label{sec:kerrBH}
It should be noted that we have the exact BH solution of a rotating BH
for Einstein's equations and we can also use it as initial data.
The Kerr BH is an exact solution of Einstein's equations in stationary
and axisymmetric spacetime\cite{Kerr1963}.
The line element of the Kerr BH in Boyer-Lindquist coordinates\cite{Boyer1966} is defined by
 \begin{equation}
  \dif s^2 = -\left(1-\frac{2Mr_{BL}}{\Sigma}\right)\dif t^2
   -\frac{4aMr_{BL}\sin^2\theta}{\Sigma}\dif t\dif\phi
   +\frac{\Sigma}{\Delta}\dif r^2_{BL} +\Sigma\dif \theta^2
   +\frac{A}{\Sigma}\sin^2\theta\dif\phi^2,
 \end{equation}
 where
 \begin{eqnarray}
  A &=& \left(r_{BL}^2+a^2\right)^2 -\Delta a^2\sin^2\theta,\\ 
  \Sigma &=& r_{BL}^2 +a^2\cos^2\theta,\\ 
  \Delta &=& r_{BL}^2 -2Mr_{BL}+a^2,
 \end{eqnarray}
 where $M$ and $a$ denote the mass and spin of BH respectively.
 $\Delta$ vanishes when the radial coordinate $r_{BL}$ is at the radius of the
 inner or outer horizon~$r_{\pm}$,
 which is a coordinate singularity.

Let us introduce a quasi-isotropic radial coordinate
in the same manner as for the Schwarzschild BH by
 \begin{eqnarray}
  r_{BL} &=& r\left(1+\frac{M+a}{2r}\right)\left(1+\frac{M-a}{2r}\right),\\
  \frac{\dif r_{BL}}{\dif r} &=& 1-\frac{M^2-a^2}{4r^2}.
 \end{eqnarray}
 Thus, the line element of the Kerr BH yields
 \begin{eqnarray}
  \dif s^2 &=& -\frac{a^2\sin^2\theta-\Delta}{\Sigma}\dif t^2
   -\frac{4aMr_{BL}\sin^2\theta}{\Sigma}\dif t\dif\phi
   +\frac{\Sigma}{r^2}\dif r^2 +\Sigma\dif\theta^2 
    +\frac{A}{\Sigma}\sin^2\theta\dif\phi^2.\nonumber\\
 \end{eqnarray}
 The spatial metric components in the quasi-isotropic coordinates
 also remain regular\cite{Brandt1996,Brandt1997}.
 One can show that the extrinsic curvature of the Kerr BH in the quasi-isotropic
 coordinates is given by
 \begin{eqnarray}
  K_{r\phi} &=& \frac{aM\left[2r_{BL}^2\left(r_{BL}^2+a^2\right)
			 +\Sigma\left(r_{BL}^2-a^2\right)\right]
  \sin^2\theta}{r\Sigma\sqrt{A\Sigma}},\\
  K_{\theta\phi} &=& \frac{-2a^3Mr_{BL}\sqrt{\Delta}\cos\theta
   \sin^3\theta}{\Sigma\sqrt{A\Sigma}},
 \end{eqnarray}
 which comes from the shift vector $\beta^{\phi}$.

\section{Apparent Horizon Finder}
\label{sec:AHF}
Now we can perform long-term dynamical simulations
containing BHs with numerical relativity.
For the sake of convenience, we usually use the apparent horizon (AH) to define the BH
and investigate the nature of BH during the evolution.
In this section, we introduce the concept of AH and derive the elliptic
PDE to determine the AH.

\subsection{Apparent Horizon}
The region of BH in an asymptotic flat spacetime is defined as the set of spacetime points
from which future-pointing null geodesics cannot reach future null infinity~\cite{Hawking1973}.
To find the BH, one can use the event horizon(EH) which is defined as the boundary of such region.
It is possible to determine the EH by the data of the numerical
simulation because in principle, one can integrate the null geodesic
equation for any spacetime points forward in time during the evolution,
\begin{eqnarray}
 \frac{\dif^2 x^{\mu}}{\dif \lambda^2} 
  +\Gamma^{\mu}_{\ \nu\rho}\frac{\dif x^{\nu}}{\dif \lambda}\frac{\dif x^{\rho}}{\dif \lambda}
  =0,
\end{eqnarray}
where $x^{\mu}$ and $\lambda$ denote the coordinates and the affine parameter.
The numerical cost to find the EH is normally high, because we need global metric data\cite{Thornburg2006}.

We define a trapped surface on the hypersurface $\Sigma$ as a smooth
closed two-dimensional surface 
on which the expansion of future-pointing null geodesics is negative.
The AH is defined as the boundary of region containing trapped surfaces
in the hypersurface and is equivalent to the marginally outer trapped surface
on which the expansion of future-pointing null geodesics vanishes\cite{kriele1997}.
The EH is outside the AH if the AH exists\cite{Hawking1973}.
We often use the AH to find the BH in the numerical simulation instead of the EH
because the AH can be locally determined and then the numerical cost is
lower compared with finding the EH.

Let us introduce the normal vector $s^i$ to the surface
and define the induced two-dimensional metric as
\begin{eqnarray}
 m_{\mu\nu} &=& \gamma_{\mu\nu} -s_{\mu}s_{\nu}
  = g_{\mu\nu}+n_{\mu}n_{\nu}-s_{\mu}s_{\nu},
\end{eqnarray}
where $\gamma_{\mu\nu}$ denotes the induced metric on the
three-dimensional hypersurface~$\Sigma$.
The null vector is described with $s^i$ and the normal vector to $\Sigma$ by
\begin{eqnarray}
 \ell^{\mu} &=& \frac{1}{\sqrt{2}}\left[s^{\mu}+n^{\mu}\right].
\end{eqnarray}
Then, the following equation should be satisfied on the AH by definition.
\begin{eqnarray}
 \Theta &=& \nabla_{\mu}\ell^{\mu} = D_is^i -K +K_{ij}s^is^j =0,\label{eq:AHeq}
\end{eqnarray}
where $\Theta$ denotes the expansion of null vector and $D_i$ denotes
the covariant derivative with respect to $\gamma_{ij}$.
\subsection{Apparent Horizon Finder}
We can find the AH during the dynamical simulation by solving Eq.~\eqref{eq:AHeq}\cite{Shibata1997,Shibata2000,Thornburg2004}.
Let us define the radius of the AH by
\begin{eqnarray}
 r=h(\theta,\phi).
\end{eqnarray}
Thus, the normal vector $s^i$ can be described with $h(\theta,\phi)$ by
\begin{eqnarray}
 \ts_i &=& \left(1,-h_{,\theta},-h_{,\phi}\right),\\
 s_i &=& C\psi^2\ts_i,\\
 C^{-2} &=& \tgam^{ij}\ts_i\ts_j,
\end{eqnarray}
where $\ts^i$ is introduced for convenience and we raise their indeces
of $s_i$ and $\ts_i$ by $\gamma^{ij}$ and $\tgam^{ij}$ respectively.
Incidentally, the divergence of the normal vector is given by
\begin{eqnarray}
 D_is^i = \frac{1}{\sqrt{\gamma}}\partial_i\sqrt{\gamma}\gamma^{ij}s_j,
\end{eqnarray}
where $\gamma$ denotes the determinant of $\gamma_{ij}$.
Therefore, we obtain the equation to determine the AH as the elliptic
PDE consisting of first and second derivatives of $h(\theta,\phi)$.
Note that because the AH equation is originally a non-linear elliptic PDE,
we change the AH equation to the flat Laplacian equation with
non-linear source term\cite{Shibata1997}, which has the advantage of
fixing the matrix with diagonal dominance mentioned in Sec.~\ref{sec:matrix}.
Specifically, we solve the following equation.
\begin{eqnarray}
 \bigtriangleup_{\theta\phi} h -\left(2-\zeta\right)h =
 h_{,\theta\theta} +\frac{\cos\theta}{\sin\theta}h_{,\theta} +\frac{1}{\sin^2\theta}h_{,\phi\phi}
  -\left(2-\zeta\right)h &=& S\left(\theta,\phi\right),\label{eq:AHequation}
\end{eqnarray}
where $\zeta$ denotes a constant to be chosen by the problem and the
source term is given by the flat laplacian term and the AH equation as
\begin{eqnarray}
 S\left(\theta,\phi\right) &=& h_{,\theta\theta} +\frac{\cos\theta}{\sin\theta}h_{,\theta} +\frac{1}{\sin^2\theta}h_{,\phi\phi}
  -\left(2-\zeta\right)h
 +\frac{h^2\psi^2}{C^3}\left[D_is^i +K_{ij}s^is^j-K\right]\nonumber\\
 &=& 2h\xi^{rr} -2h\tgam^{r\theta}h_{,\theta} -2h\tgam^{r\phi}h_{,\phi}
  +h^2\cot\theta \tgam^{\theta r} -h^2\cot\theta\ \tgam^{\theta\phi}h_{,\phi}\nonumber\\
 & & -h^2\xi^{\theta\theta}h_{,\theta\theta}
  -h^2\xi^{\phi\phi}h_{,\phi\phi}
  +\zeta h \nonumber\\
 & & +\frac{1-C^2}{C^2}\left[
  2h \ts^{r}+h^2\cot\theta \ts^{\theta}
  -h^2\tgam^{\theta\theta}h_{,\theta\theta}
  -h^2\tgam^{\phi\phi}h_{,\phi\phi} \right]\nonumber\\
 & & +\frac{h^2C_{,i}\ts^i}{C^{3}} +\frac{4h^2\psi_{,i}\ts^i}{C^2\psi} -\frac{h^2\tGam^j\ts_j}{C^{2}}
  -\frac{2h^2\tgam^{\theta\phi}h_{,\theta\phi}}{C^{2}}\nonumber\\
 & & +\frac{h^2\psi^2}{C}\tA_{ij}\ts^i\ts^j -\frac{2h^2\psi^2}{3C^3}K,
\end{eqnarray}
where $\xi^{ij}\equiv\tgam^{ij}-\eta^{ij}$.

\subsection{Mass and Spin of Black Hole}
The area of the AH is defined by
\begin{eqnarray}
 \mathcal{A}_{AH} &=& \int_{S}\sqrt{\det(g_{\mu\nu})}\,\dif S,
\end{eqnarray}
where $S$ denotes the surface of AH.
We also compute the quantities related to the AH,
the polar and equatorial circumfential length($\mathcal{C}_p, \mathcal{C}_e$).
\begin{eqnarray}
 \mathcal{C}_{p} &=& \int_{0}^{\pi}\dif\theta
  \sqrt{g_{rr}h_{,\theta}^2 +g_{r\theta}h_{,\theta}
  +g_{\theta\theta}},\\
 \mathcal{C}_{e} &=& \int_{0}^{2\pi}\dif\phi
  \sqrt{g_{rr}h_{,\phi}^2 +g_{r\phi}h_{,\phi}
  +g_{\phi\phi}}.
\end{eqnarray}
If the BH relaxes to a stationary state during the evolution, the BH
would be the Kerr BH because of no-hair theorem.
The quantities related to the AH of the Kerr BH can be obtained by
\begin{eqnarray}
 \mathcal{A}_{AH} &=& 8\pi M_{BH}^2\left(1+\sqrt{1-a^2}\right),\\
 \mathcal{C}_{e} &=& 4\pi M_{BH},\\
 \frac{\mathcal{C}_{p}}{\mathcal{C}_{e}} &=& \frac{\sqrt{2r_{+}}}{\pi}E(\frac{a^2}{2r_{+}}),
\end{eqnarray}
where $M_{BH}$, $a$ and $r_{+}$ denote the mass, spin and outer horizon radius
defined by $r_{+}=1+\sqrt{1-a^2}$ and $E(z)$ denotes an elliptic
integral defined by
\begin{eqnarray}
 E(z) &=& \int_0^{\pi/2}\sqrt{1-z\sin^2\theta}\dif \theta.
\end{eqnarray}

\section{Numerical Methods for solving elliptic PDEs}
\label{sec:numerical_methods}
Constructing the initial data for numerical relativity
is, in general, equivalent to solving the elliptic PDEs \eqref{eq:ham_const_ivp}
and \eqref{eq:mom_const_ivp} with appropriate conditions.
In order to solve a binary problem with high accuracy,
the spectral method should be the standard method for solving an elliptic
PDEs. In fact, there are useful open source codes, for example,
TwoPuncture\cite{Ansorg2004ds,Ansorg2006gd} and LORENE\cite{Grandclement2007sb}.

Futhermore, we have to solve another elliptic PDE
to find the BH in simulations within numerical relativity
as described in Sec.~\ref{sec:AHF}.
In this case, fast methods to solve the ellitptic PDE are preferred. 
Because there are many elliptic PDE solvers, the method has to be chosen according to the specific purpose.
In this section, we introduce some classical numerical methods for beginners.
It would also be the basis for Multi-Grid method mentioned in~\ref{sec:MultiGrid}.

\subsection{Discretization}
We should discretize our physical space to the computational grid space by finite difference method
because we cannot take continuum fields into account on the computer.
Consider first one-dimensional problems for simplicity, and
introduce the grid interval $\Delta x$.
Taylor expansion of a field $Q(x)$ is given by
\begin{eqnarray}
 Q(x+\Delta x) &=& Q(x) +\Delta x\frac{\partial Q}{\partial x}
  +\frac{\Delta x^2}{2}\frac{\partial^2 Q}{\partial x^2}
  +\frac{\Delta x^3}{6}\frac{\partial^3 Q}{\partial x^3}
  +\mathcal{O}(\Delta x^4),\\
 Q(x-\Delta x) &=& Q(x) -\Delta x\frac{\partial Q}{\partial x}
  +\frac{\Delta x^2}{2}\frac{\partial^2 Q}{\partial x^2}
  -\frac{\Delta x^3}{6}\frac{\partial^3 Q}{\partial x^3}
  +\mathcal{O}(\Delta x^4).
\end{eqnarray}
Thus, the derivative of the field $Q(x)$ can be expressed as
\begin{eqnarray}
 \frac{\partial Q}{\partial x}(x) &=& \left\{
  \begin{array}{cll}
   \displaystyle\frac{Q_{j+1}-Q_{j}}{\Delta x} +\mathcal{O}(\Delta x) &,& {\rm forward\, difference},\\[4mm]
   \displaystyle\frac{Q_{j}-Q_{j-1}}{\Delta x} +\mathcal{O}(\Delta x) &,& {\rm backward\, difference},
  \end{array}\right.
\end{eqnarray}
where $Q_{j+1},Q_j$ and $Q_{j-1}$ denote $Q(x+\Delta x), Q(x)$ and
 $Q(x-\Delta x)$ respectively and
 both accuracies of the forward and backward difference method for
derivatives are $\mathcal{O}(\Delta x)$.  In addition, the central
difference method whose accuracy is $\mathcal{O}(\Delta x^2)$ 
can be defined by both Taylor expansions as
\begin{eqnarray}
 \frac{\partial Q}{\partial x}(x) &=&\frac{Q_{j+1}-Q_{j-1}}{\Delta x} +\mathcal{O}(\Delta x^2).
\end{eqnarray}
Similarly, the second-order derivative of $Q(x)$ is written by
\begin{eqnarray}
 \frac{\partial^2 Q}{\partial x^2}(x) &=& \frac{Q_{j+1}-2Q_{j}+Q_{j-1}}{\Delta x^2}
  +\mathcal{O}(\Delta x^2).
\end{eqnarray}

One can increase accuracy of the calculation by using many points.
For example, using five values
$Q(x+2\Delta x), Q(x+\Delta x),Q(x), Q(x-\Delta x)$
and $Q(x-2\Delta x)$ around $x$,
the fourth-order accuracy scheme are defined by
\begin{eqnarray}
 \frac{\partial Q}{\partial x}(x) &=&
  \frac{-Q_{j+2} +8Q_{j+1} -8Q_{j-1} +Q_{j-2}}{12\Delta x}
  +\mathcal{O}(\Delta x^4),\\
 \frac{\partial^2 Q}{\partial x^2}(x) &=&
  \frac{-Q_{j+2} +16Q_{j+1} -30Q_{j} +16Q_{j-1} -Q_{j-2}}{12\Delta x^2}
  +\mathcal{O}(\Delta x^4),
\end{eqnarray}
noting that higher accuracy scheme can also be defined.
We also note that we can discretize our space in more than two dimensions
in the same way.

\subsection{Relaxation Method}
Hereafter, let us focus on Poisson equations $(\bigtriangleup\psi=S)$
with a field $\psi$ and a source $S$ as elliptic PDEs. These are a sufficiently general and complex class of problems that they embody 
all necessary elements to solve Poisson equation for constructing initial data for numerical relativity or finding an apparent horizon of BH.
We explain how to solve general elliptic PDEs in \ref{app:LDUdecomp}.
One of the simple methods to solve elliptic PDEs, so-called relaxation
method\cite{Press2007,Varga2009,Hageman2012}, is described in this section.
Let us introduce a virtual time $\tau$ to solve an elliptic PDE
and our equation of elliptic type can be transformed to the equation of parabolic type as
\begin{eqnarray}
 \frac{\partial\psi}{\partial \tau} = \bigtriangleup\psi -S,
\end{eqnarray}
which denotes the original Poisson equation after $\psi$ relaxes by iteration.
We adopt Cartesian coordinates in three-dimensional spaces and
discretize the Poisson equation with second-order accuracy as
\begin{eqnarray}
 \frac{\psi^{n+1}_{j,k,l}-\psi^{n}_{j,k,l}}{\Delta\tau}
  &=&\frac{\psi^{n}_{j+1,k,l}-2\psi^{n}_{j,k,l}+\psi^{n}_{j-1,k,l}}{\Delta x^2}
  +\frac{\psi^{n}_{j,k+1,l}-2\psi^{n}_{j,k,l}+\psi^{n}_{j,k-1,l}}{\Delta y^2}\nonumber\\
 &+&\frac{\psi^{n}_{j,k,l+1}-2\psi^{n}_{j,k,l}+\psi^{n}_{j,k,l-1}}{\Delta z^2}
  -S_{j,k,l},
\end{eqnarray}
where the superscript $n$ denotes the label of virtual time and
the subscript $j,k$ and $l$ denote labels of $x-,y-$ and $z-$direction, respectively.
Therefore, the field in the next step of the iteration
is determined by
\begin{eqnarray}
 \psi^{n+1}_{j,k,l} &=&
  \left[1-2\Delta\tau \left(\frac{1}{\Delta x^2}+\frac{1}{\Delta y^2}+\frac{1}{\Delta z^2} \right)
   \right]\psi^{n}_{j,k,l}
  -\Delta\tau\, S_{j,k,l}
  \nonumber\\
 &+&\Delta\tau\left[\frac{\psi^{n}_{j+1,k,l}+\psi^{n}_{j-1,k,l}}{\Delta x^2}
	       +\frac{\psi^{n}_{j,k+1,l}+\psi^{n}_{j,k-1,l}}{\Delta y^2}
	       +\frac{\psi^{n}_{j,k,l+1}+\psi^{n}_{j,k,l-1}}{\Delta z^2} \right].
\end{eqnarray}
We continue to update the field $\psi$ by the above expression until $\psi$ relaxes
and obtain the solution of the Poisson equation~($\bigtriangleup\psi=S$).

\subsubsection{Jacobi Method}
\label{sec:method_jacobi}
In order to discuss method in practice, we consider one-dimensional
Poisson equation and discretize it as
\begin{eqnarray}
 \bigtriangleup\psi=\frac{\dif^2 \psi}{\dif x^2} &=& S,\nonumber\\
 \frac{\psi_{j+1}-2\psi_{j}+\psi_{j-1}}{\Delta x^2}
  &=& S_{j}.\label{eq:discretize_jacobi}
\end{eqnarray}
We consider Eq.~(\ref{eq:discretize_jacobi}) as the equation to
determine $\psi_{j}$,
\begin{eqnarray}
 \psi_{j}^{n+1,\mJ} &=& \frac{1}{2}\left[\psi_{j+1}^{n} +\psi_{j-1}^{n} -\Delta x^2S_{j}^{n}\right],\label{eq:method_jacobi}
\end{eqnarray}
where the superscript $n$ denotes the label of time step and we attach the label
$\mJ$ on the field of next time step in Jacobi's method.
Thus, we repeat updating the field until $\psi$ converges.
In other words, the flowchart of Jacobi method is as follows.
\begin{itemize}
 \item[1.] Give a trial field $\psi^n$.
 \item[2.] We obtain a new field $\psi^{n+1}$ by Eq.~(\ref{eq:method_jacobi}).
 \item[3.] Set the obtained field to a new trial field.
 \item[4.] Repeat these steps~(1.-3.) until the change of $\psi$ is within a 
       numerical error.
\end{itemize}

In addition, it is easy to extend to the three-dimensional Poisson equation as
\begin{eqnarray}
 \psi_{j,k,l}^{n+1,\mJ} &=& \frac{1}{6}
  \left[\psi_{j+1,k,l}^{n} +\psi_{j-1,k,l}^{n}
   +\psi_{j,k+1,l}^{n} +\psi_{j,k-1,l}^{n}
   +\psi_{j,k,l+1}^{n} +\psi_{j,k,l-1}^{n}\right]\nonumber\\
& &   -\Delta h^2 S_{j,k,l}^{n},
\end{eqnarray}
where we define $\Delta h\equiv\Delta x =\Delta y=\Delta z$ for simplicity.

\subsubsection{Matrix expression}
\label{sec:matrix}
Discretized elliptic PDEs can be expressed by matrices and vectors.
Once we describe the elliptic PDE via a matrix expression, the problem involves solving the
inverse of the matrix.
For example, a matrix expression for Jacobi method is given as follows.
We introduce a solution vector $\psi_I$ and source vector $b_I$.
Then, Eq.~(\ref{eq:discretize_jacobi}) can be expressed as
\begin{eqnarray}
 \left[
  \begin{array}{ccccccc}
   A_{00}&A_{01}  & A_{02}& A_{03} &\cdots& A_{0 N-1} & A_{0N}\\
   1     & -2     & 1     & 0 &\cdots& 0 & 0\\
   0     & 1      & -2    & 1 &\cdots& 0 & 0\\
   \vdots& \vdots & \vdots& \vdots &\ddots& \vdots& \vdots\\
   0&0  & 0 & 0 &\cdots& -2& 1\\
   A_{N0}&A_{N1}  & A_{N2} & A_{N3} &\cdots& A_{N N-1}& A_{NN}
  \end{array}
 \right]
 \left[
  \begin{array}{c}
   \psi_0\\
   \psi_1\\
   \psi_2\\
   \vdots\\
   \psi_{N-1}\\
   \psi_N
  \end{array}
 \right]=
 \left[
  \begin{array}{c}
   b_0\\
   \Delta x^2S_1\\
   \Delta x^2S_2\\
   \vdots\\
   \Delta x^2S_{N-1}\\
   b_N
  \end{array}
 \right],
\end{eqnarray}
where $A_{IJ}$ is the coefficient matrix corresponding to the Laplacian operator
and the first and last rows of $A_{IJ}$ denote boundary conditions to be determined
by physics.
It is formally expressed by $A_{IJ}\psi_{J}=b_{I}$ and the problem
leads to solving the inverse of coefficient matrix as $\psi_{J}=A_{IJ}^{-1}b_{I}$.
There are many methods to numerically solve the inverse of matrices.
In general, Jacobi method for an arbitrary coefficient matrix is expressed as
\begin{eqnarray}
 A_{II}\psi_{I} +\sum_{I\neq J}A_{IJ}\psi_{J} &=& b_{I},\nonumber\\
 \psi_{I}^{n+1, \mJ} &=& \frac{1}{A_{II}}\left(b_{I} -\sum_{I\neq J}A_{IJ}\psi_{J}^{n}\right).
\end{eqnarray}
We note that other elliptic PDEs may not be solvable by the Jacobi method
because the iteration is not always stable; this can be shown by the von Neumann numerical stability analysis.
However, the Poisson equation is fortunately stable, which is equivalent to that
the matrix is diagonally dominant.

\subsubsection{Gauss-Seidel Method}
\label{sec:method_gs}
In iterative methods to solve Poisson equations as Jacobi method,
it depends on a trial field $\psi^{n}$ how fast we obtain solutions.
We usually expect that it becomes better solution as iteration step goes forward.
In order to obtain a closer trial field to the solution,
we should actively use updated values.
Thus, by Gauss-Seidel method, we can determine a next trial field as
\begin{eqnarray}
 \psi_{j}^{n+1,\mGS} &=& \frac{1}{2}\left[\psi_{j+1}^{n} +\psi_{j-1}^{n+1,\mGS} -\Delta x^2S_{j}^{n}\right],
\end{eqnarray}
where $\mGS$ denotes that the field is determined by Gauss-Seidel method.
In addition, the matrix expression for Gauss-Seidel method is given by
\begin{eqnarray}
 A_{II}\psi_{I} +\sum_{I< J}A_{IJ}\psi_{J} +\sum_{I> J}A_{IJ}\psi_{J}=b_{I},\nonumber\\
 \psi_{I}^{n+1,\mGS} = \frac{1}{A_{II}}\left(b_{I} -\sum_{I> J}A_{IJ}\psi_{J}^{n} -\sum_{I< J}A_{IJ}\psi_{J}^{n+1}\right).
\end{eqnarray}

\subsubsection{SOR Method}
\label{sec:method_sor}
It turns out that the Poisson equation is faster to solve with Gauss-Seidel
method than with the Jacobi method, as shown later.
Although the speed with which the numerical solution converges depends on the trial
field in iterative methods, the Gauss-Seidel method gives
a ``better'' field than Jacobi's in that respect.
Thus, it is possible to accelerate convergence by
specifying trial guess of the field more aggressively.
This method is called Successive Over-Relaxation(SOR) method,
which is defined by
\begin{eqnarray}
 \psi^{n+1,\mS}_{j} = \psi^{n}_{j} +\omega\left(\psi_{j}^{n+1,\mGS} -\psi_{j}^{n}\right),
\end{eqnarray}
where the superscript $\mS$ denotes the label of SOR method and
$\omega$ denotes an acceleration parameter whose range is defined in
$1\leq\omega<2$ by the stability analysis.
When we set the acceleration parameter as unity, SOR method is
identical to the Gauss-Seidel method by definition.

\section{Results}
In this section, we introduce sample codes
to solve Poisson equations using different methods.
These codes are sufficiently general that they can be applied to other problems in physics, provided one slightly 
changes the source term and boundary conditions.

\subsection{Code Tests}
As tests for our codes, we use the following analytical solutions.
We show numerical results of Poisson equations with simple linear source and
sufficiently non-linear source. In addition, the code to find the AH of the Kerr BH is also shown as an example.
Some sample codes parallelized with OpenMP are also available~\ref{app:samplecode}.

\subsubsection{Linear source}
\label{sec:test_linear}
\begin{figure}[ht]
\begin{tabular}{cc}
\psfig{file=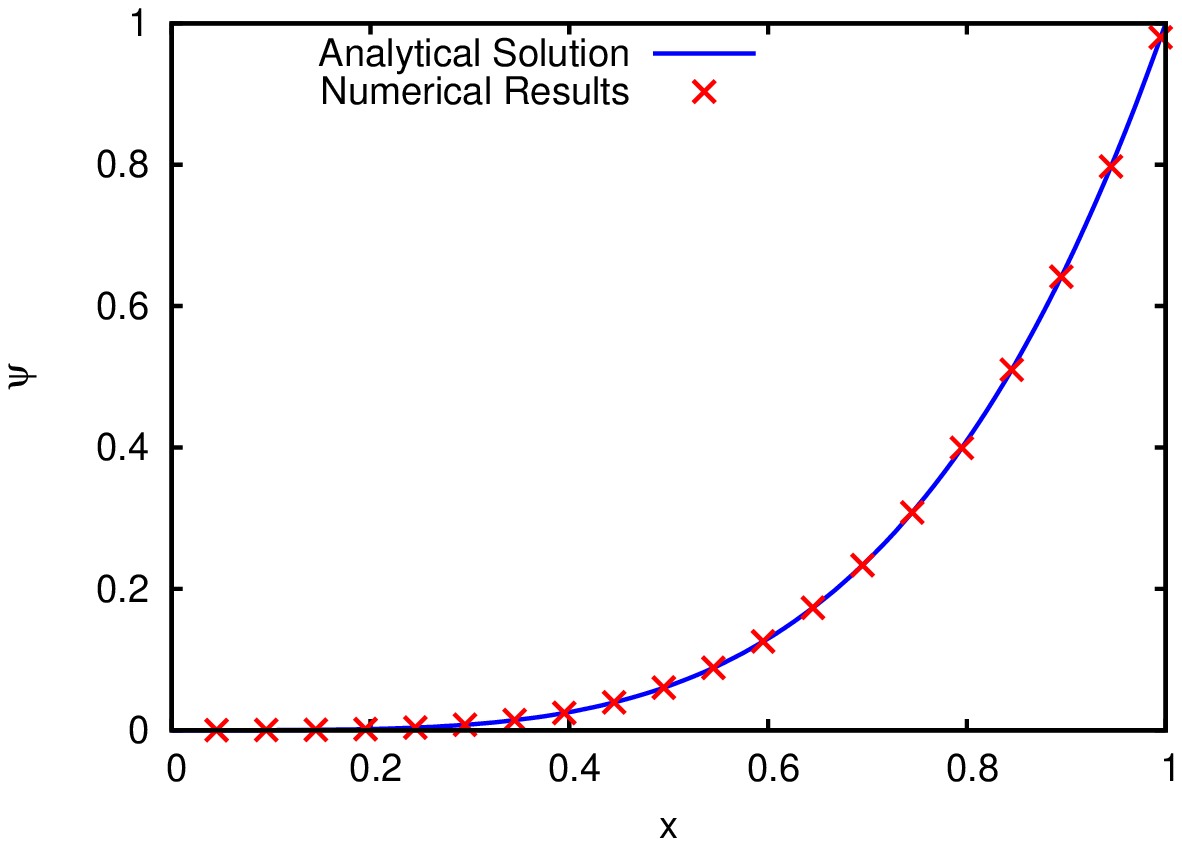,width=6.cm}
 &
\psfig{file=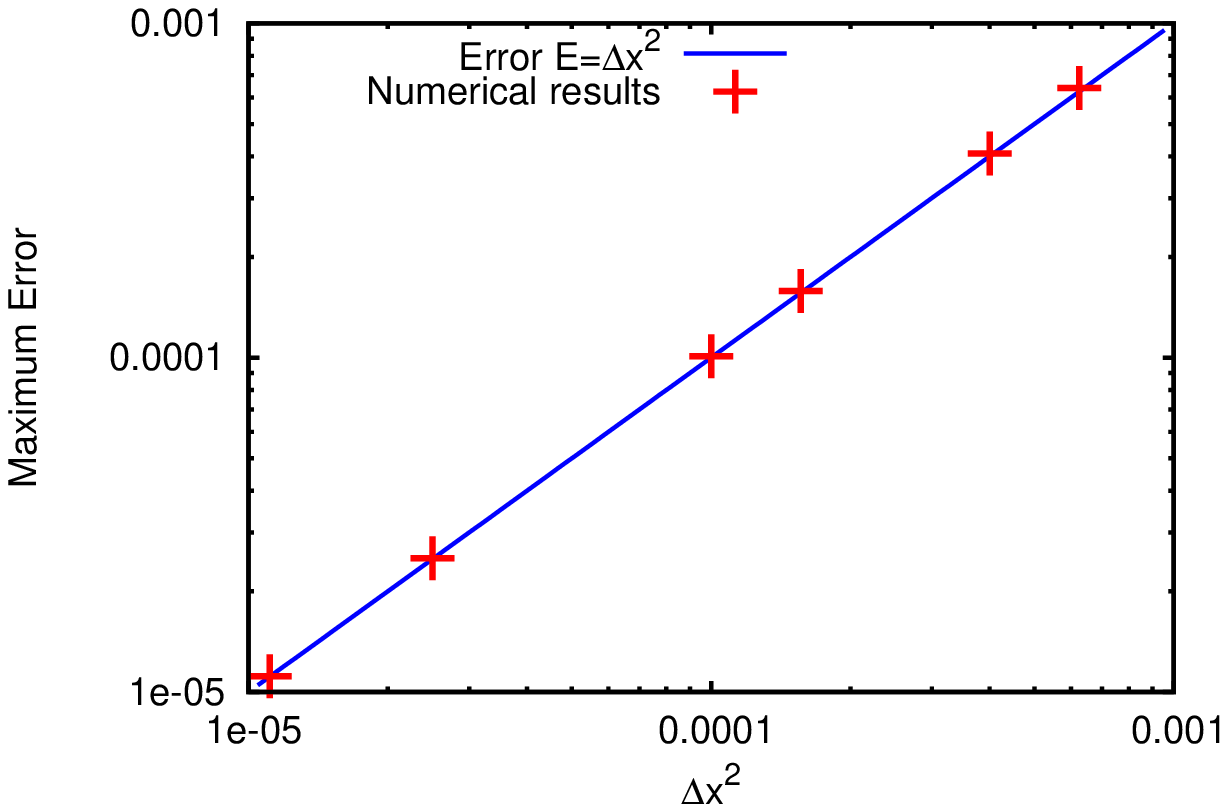,width=6.cm}\\
 (a)& (b)\\
\multicolumn{2}{c}{\psfig{file=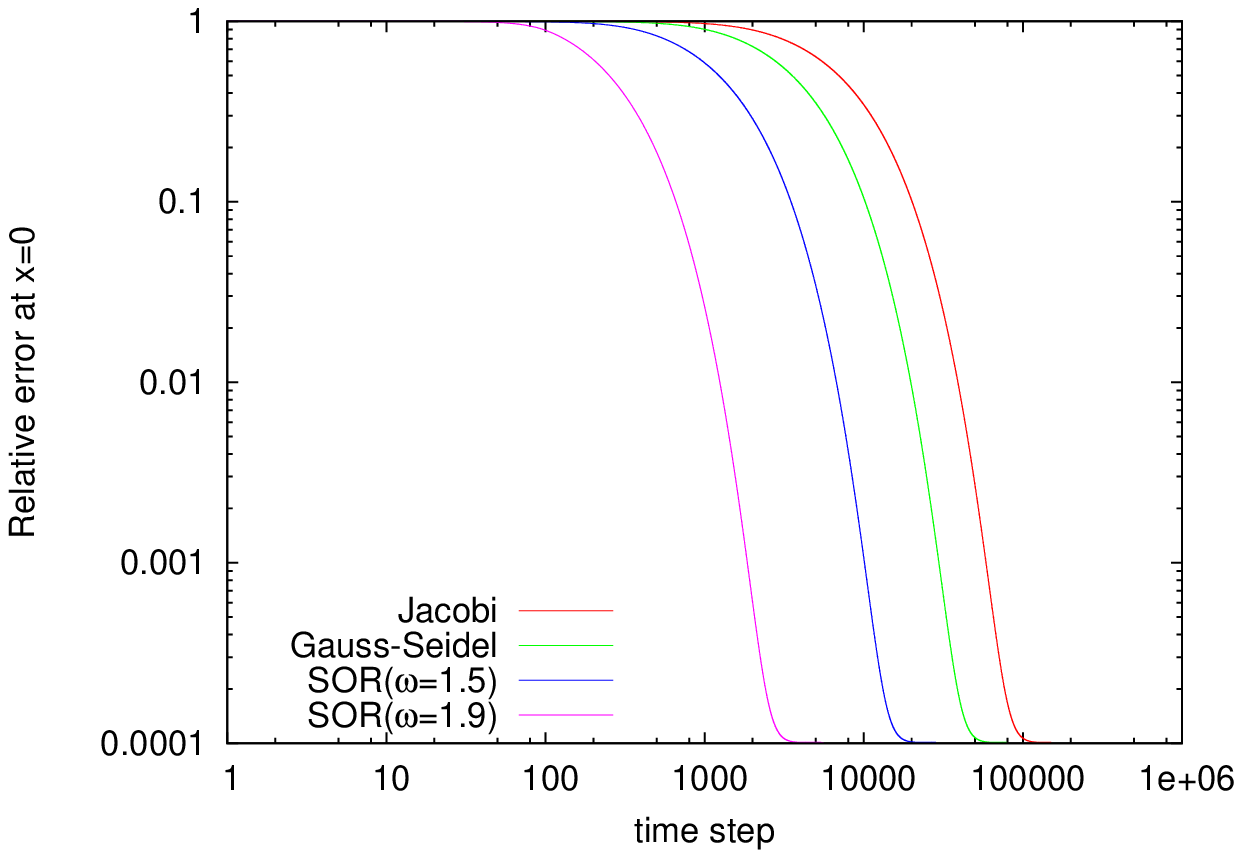,width=8.cm}}\\
\multicolumn{2}{c}{(c)}
\end{tabular}
 \vspace*{6pt}
 \caption{(a) Numerical solution by Jacobi method compared to the
 analytical solution.
 (b) The convergence test by the maximum relative error between analytical
 solution and numerical result as a function of the resolution. 
 (c) The difference of iterative time steps needed to converge among
 methods to solve the Poisson equation. Vertical axis denotes the
 relative error between the numerically obtained solution and analytical
 solution at $x=0$.
 \label{fig:src1}}
\end{figure}

Let us consider simple source term for the Poisson equation as
\begin{eqnarray}
 \bigtriangleup\psi = \frac{\dif^2 \psi}{\dif x^2} = 12 x^2.\label{eq:analytic_src1}
\end{eqnarray}
In numerical computation, we set the range 
as $0 \leq x \leq 1$ and boundary conditions by
\begin{eqnarray}
 \frac{\dif \psi}{\dif x}\biggr|_{x=0}= 0&,& {\rm Neumann\, B.C.},\nonumber\\
 \psi\bigr|_{x=1}= 1&,& {\rm Dirichlet\, B.C.}.\label{eq:bc_src1}
\end{eqnarray}
Then, we obtain the analytical solution by integrating
Eq.~(\ref{eq:analytic_src1}) twice with boundary conditions~(\ref{eq:bc_src1}),
\begin{eqnarray}
 \psi(x) &=& x^4.
\end{eqnarray}
Arbitrary initial guess for the solution can be given and
we set $\psi(x)=1$ at initial for those boundary conditions.
We set the resolution of the computational grid as $\Delta x=1/100$.
Fig.~\ref{fig:src1}~(a) shows the numerical solution by Jacobi method
as compared to the analytical solution.
We note that the accuracy of the numerical result depends on the
computational resolution and how the accuracy increase with resolution depends on
the scheme of discretization; Fig.~(b) is compatible with second-order accuracy.
We compare Poisson solvers in Fig.~\ref{fig:src1}~(c)
by the time steps needed to obtain the solution.
Curves show the difference of methods and we choose the Jacobi
method, Gauss-Seidel method,
SOR methods with $\omega=1.5$ and $\omega=1.9$.
The SOR method gives the solution about 10-100 times faster than Jacobi method,
and depends on the acceleration parameter $\omega$.

\subsubsection{Non-linear source}
\label{sec:test_nonlinear}
Next, let us consider a weak gravitational
field, namely Newtonian gravitational source.
A gravitational potential $\Phi$ can be determined by
the Poisson equation,
\begin{eqnarray}
 \bigtriangleup\Phi &=& -4\pi\rho,
\end{eqnarray}
where we omit the Newton constant by using $G=1$ units.
Suppose gravitational sources are distributed with spherical symmetry as
\begin{eqnarray}
 \rho(r) &=& \left\{
	      \begin{array}{cr}
	       \rho_0\left(1-r^2\right),\ \ \  & r<1,\\
	       0, & r\geq 1,
	      \end{array}
		  \right.\label{eq:source_grav}
\end{eqnarray}
where $\rho_0$ is a constant.
Corresponding Poisson equation is rewritten by 
 \begin{eqnarray}
  \bigtriangleup\Phi = \frac{1}{r^2}\frac{\partial}{\partial r}\left[r^2\frac{\partial}{\partial r}\right]\Phi
   +\frac{1}{r^2\sin\theta}\left[\sin\theta\frac{\partial}{\partial\theta}\right]\Phi
   +\frac{1}{r^2\sin^2\theta}\frac{\partial^2}{\partial\phi^2}\Phi &=& -4\pi \rho\nonumber\\
  \frac{1}{r^2}\frac{\partial}{\partial r}\left[r^2\frac{\partial}{\partial r}\right]\Phi 
   &=&-4\pi\rho.\nonumber\\
 \end{eqnarray}
 Thus, we obtain the analytical solution of the source~(\ref{eq:source_grav})
 by solving the equations separately as the region ($r>1$) with the boundary
 condition $\Phi\rightarrow 0$ at infinity and ($r\leq 1$) with the regularity condition at the origin.
\begin{eqnarray}
  \Phi(r) = 
   \left\{
    \begin{array}{cr}
     \displaystyle
      \pi\rho_0\left[\frac{r^4}{5}-\frac{2r^2}{3}+1\right],\ \ \ & r\leq 1,\\
     \displaystyle \frac{8\pi\rho_0}{15 r}, & r > 1.
    \end{array}
   \right.
\end{eqnarray}
We consider this analytical solution to test our code.
The Poisson equation with spherical symmetry can be regarded as one
dimentional Poisson equation with the non-linear source in our method,
\begin{eqnarray}
 \frac{\partial^2\Phi}{\partial r^2}
   =-4\pi\rho -\frac{\partial\Phi}{\partial r},\label{eq:nlPoisson}
\end{eqnarray}
whose range to be considered as $0\leq x\leq 10$
and boundary conditions are set as
\begin{eqnarray}
 \frac{\dif \Phi}{\dif r}\biggr|_{r=0}= 0&,& {\rm Neumann\, B.C.},\nonumber\\
 \frac{\dif \left(r\Phi\right)}{\dif r}\biggr|_{r=10}= 0&,& {\rm Robin\, B.C.},\label{eq:bc_src2}
\end{eqnarray}
where the Robin boundary condition is chosen because we expect
$\Phi\rightarrow r^{-1}$ at large distance.
\begin{figure}[ht]
\begin{tabular}{cc}
 \psfig{file=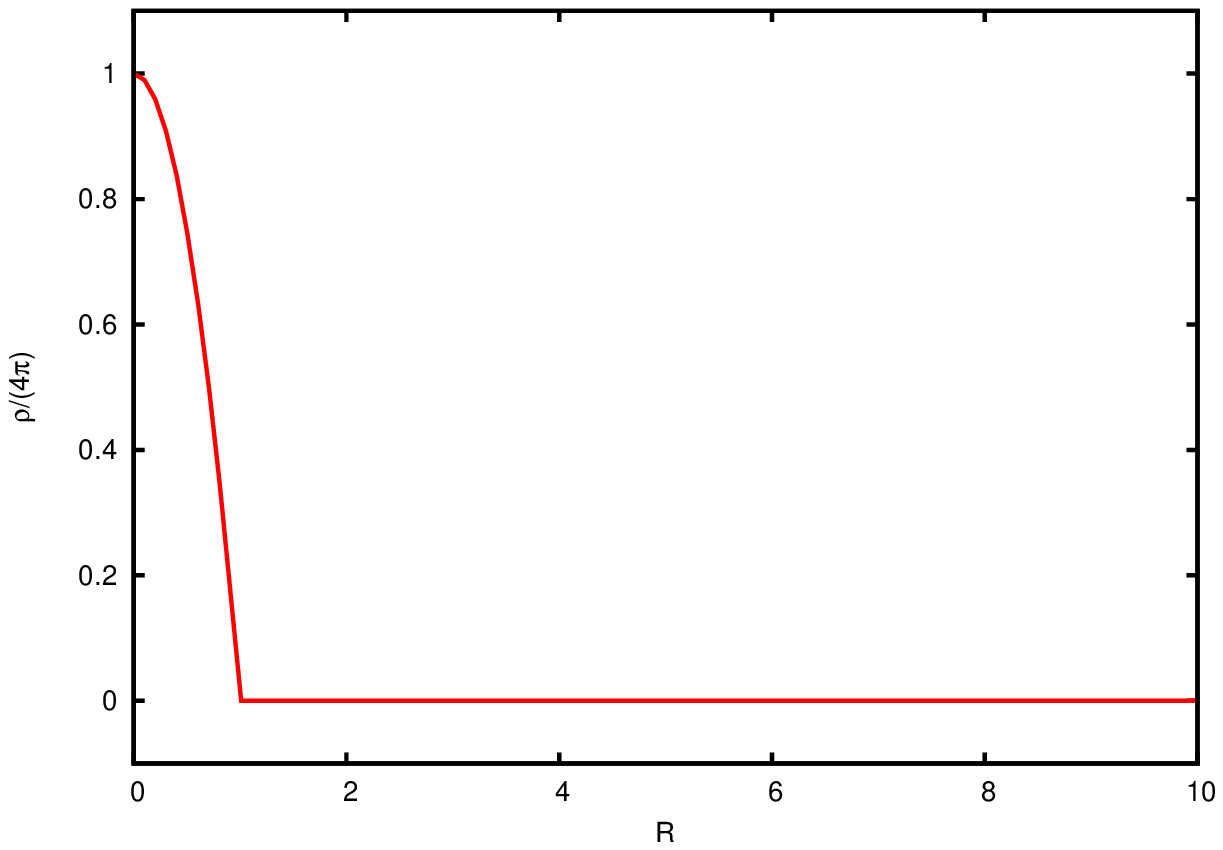,width=6.cm}
 &
 \psfig{file=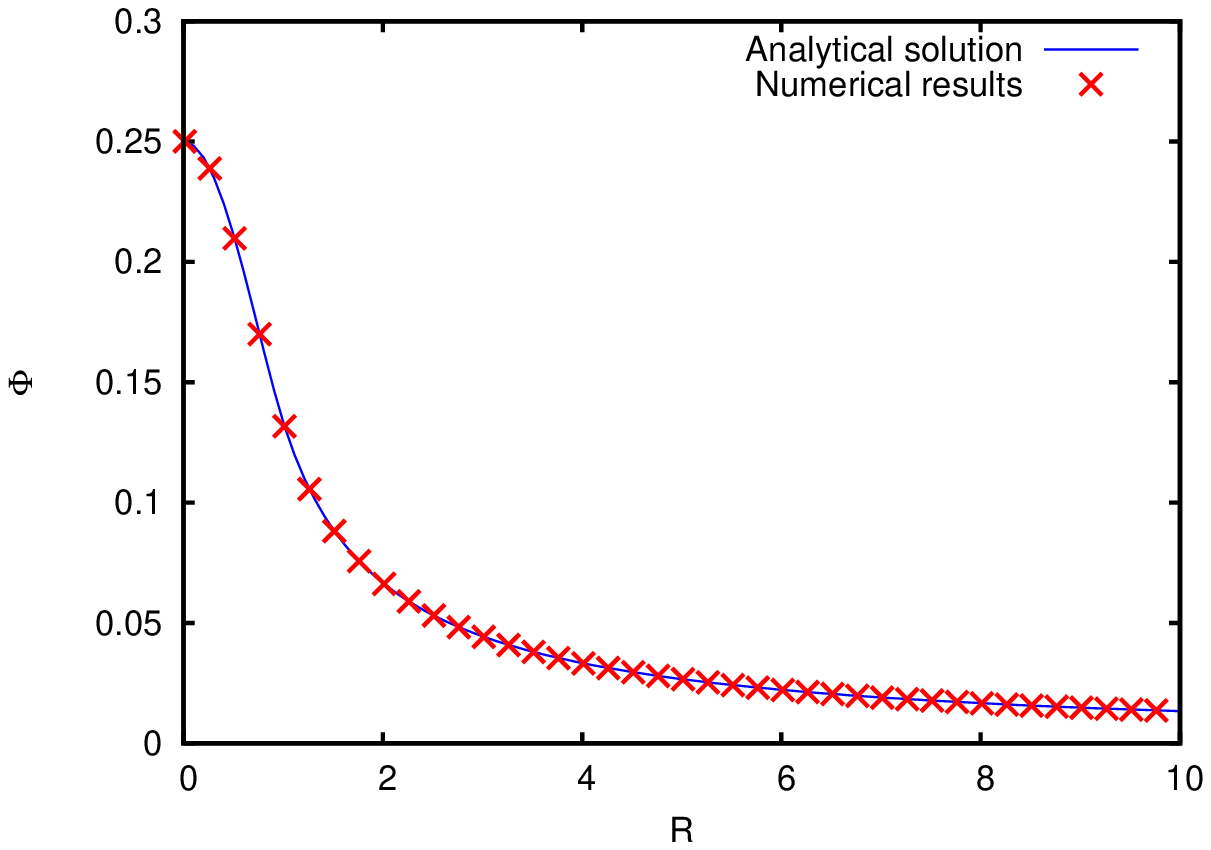,width=6.cm}\\
 (a)& (b)
\end{tabular}
 \vspace*{2pt}
 \caption{(a) Distribution of the gravitational source as a function of $R$.
 (b) Numerical solution of the non-linear source by SOR method
 compared to analytical solution.}
 \label{fig:src2}
\end{figure}
Fig.~\ref{fig:src2}~(a) shows the source distribution and
(b) shows the numerical result by solving the Eq.~(\ref{eq:nlPoisson}).
The result is obtained with 400 grid points but shown with only 40 points.
\subsubsection{Apparent Horizon of Kerr Black Hole}
\label{sec:test_KBH}
Let us apply our code for Poisson solver to solving the AH of Kerr BH.
The AH equation~(\ref{eq:AHequation}) should be reduced to simpler
equation with axisymmetry $\partial_{\phi}=0$.

The normal vector $s_i$ is defined with axisymmetry and
the normalization $C$ is determined by the Kerr metric as
\begin{eqnarray}
 \bs_i &=& \left[1, -h_{,\theta}, 0\right],\ \  s_i = C\bs_i,\\
 C^{-2} &=& \gamma^{ij}\bs_i\bs_j = \gamma^{rr}+\gamma^{\theta\theta}h_{,\theta}^2.
\end{eqnarray}
To be concrete, we note that the non-trivial part of the AH equation with axisymmetry
in isotropic coordinates can be written by
\begin{eqnarray}
 D_is^i &=&  \frac{1}{\sqrt{\gamma}}\partial_i\sqrt{\gamma}\gamma^{ij}s_j\nonumber\\
 &=& -\frac{C}{h}\gamma^{rr}
 +\frac{C}{2\Sigma}\left(\Sigma_{,r}\gamma^{rr} -\Sigma_{,\theta}\gamma^{\theta\theta}h_{,\theta}\right)
 +\frac{C}{2A}\left(A_{,r}\gamma^{rr} -A_{,\theta}\gamma^{\theta\theta}h_{,\theta}\right)\nonumber\\
 & &
  -C\gamma^{\theta\theta}\cot\theta h_{,\theta}
  +C\gamma^{rr}_{,r}
  -Ch_{,\theta}\gamma^{\theta\theta}_{,\theta}
  -C\gamma^{\theta\theta}h_{,\theta\theta}\nonumber\\
 & & -\frac{C^3}{2}\gamma^{rr}\left[\gamma^{rr}_{,r}+\gamma^{\theta\theta}_{,r}h_{,\theta}^2\right]
  +\frac{C^3}{2}\gamma^{\theta\theta}h_{,\theta}
  \left[\gamma^{rr}_{,\theta}+\gamma^{\theta\theta}_{,\theta}h_{,\theta}^2 +2\gamma^{\theta\theta}h_{,\theta}h_{,\theta\theta}\right],
\end{eqnarray}
where
\begin{eqnarray}
 \frac{\dif r_{BL}}{\dif r}
  &=& 1 -\frac{M^2-a^2}{4r^2},\ \ 
 C_{,i} =
  -\frac{C^3}{2}\left[\gamma^{rr}_{,i}+\gamma^{\theta\theta}_{,i}h_{,\theta}^2
		 +2\gamma^{\theta\theta}h_{,\theta}h_{,\theta i}\right],\nonumber\\
 \Sigma_{,r} &=& 2r_{BL}\frac{\dif r_{BL}}{\dif r},\ \ 
 \Sigma_{,\theta} = -2a^2\cos\theta\sin\theta,\ \ 
 \Delta_{,r} = 2\left(r_{BL}-M\right)\frac{\dif r_{BL}}{\dif r},\nonumber\\
 A_{,r} &=& 4\left(r_{BL}^2+a^2\right)r_{BL}\frac{\dif r_{BL}}{\dif r}
  -\Delta_{,r}a^2\sin^2\theta,\ \ 
  A_{,\theta} = -2\Delta a^2\sin\theta\cos\theta,\nonumber\\
\gamma^{rr}_{,r} 
 &=& \frac{2r\Sigma-r^2\Sigma_{,r}}{\Sigma^2},\ \ 
\gamma^{rr}_{,\theta} 
= -\frac{r^2\Sigma_{,\theta}}{\Sigma^2},\ \ 
\gamma^{\theta\theta}_{,r} 
= -\frac{\Sigma_{,r}}{\Sigma^2},\ \ 
\gamma^{\theta\theta}_{,\theta} 
= -\frac{\Sigma_{,\theta}}{\Sigma^2}.
\end{eqnarray}

On the other hand, we can also solve the AH in Boyer-Lindquist
coordinates, only to change the following part.
\begin{eqnarray}
 D_is^i 
 &=&
  \frac{C}{2\Sigma}\left(\Sigma_{,r}\gamma^{rr} -\Sigma_{,\theta}\gamma^{\theta\theta}h_{,\theta}\right)
 +\frac{C}{2A}\left(A_{,r}\gamma^{rr} -A_{,\theta}\gamma^{\theta\theta}h_{,\theta}\right)
 -\frac{C}{2\Delta}\Delta_{,r}\gamma^{rr}
 \nonumber\\
 & &
  -C\gamma^{\theta\theta}\cot\theta h_{,\theta}
  +C\gamma^{rr}_{,r}
  -Ch_{,\theta}\gamma^{\theta\theta}_{,\theta}
  -C\gamma^{\theta\theta}h_{,\theta\theta}\nonumber\\
 & & -\frac{C^3}{2}\gamma^{rr}\left[\gamma^{rr}_{,r}+\gamma^{\theta\theta}_{,r}h_{,\theta}^2\right]
  +\frac{C^3}{2}\gamma^{\theta\theta}h_{,\theta}
  \left[\gamma^{rr}_{,\theta}+\gamma^{\theta\theta}_{,\theta}h_{,\theta}^2 +2\gamma^{\theta\theta}h_{,\theta}h_{,\theta\theta}\right].
\end{eqnarray}

\begin{figure}[ht]
\begin{tabular}{cc}
\psfig{file=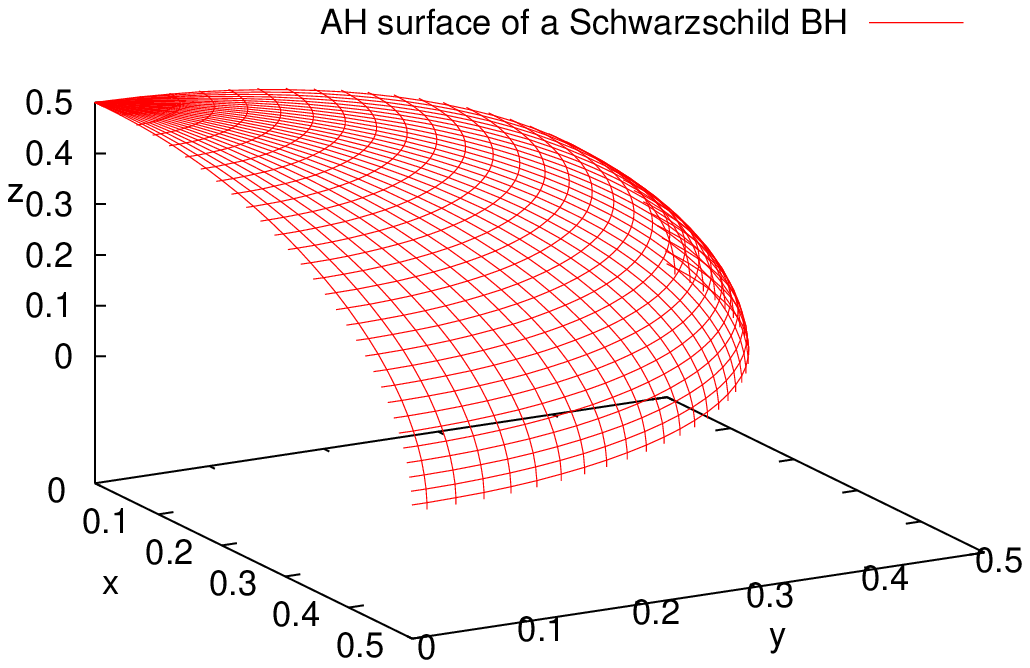,width=6.cm}&
\psfig{file=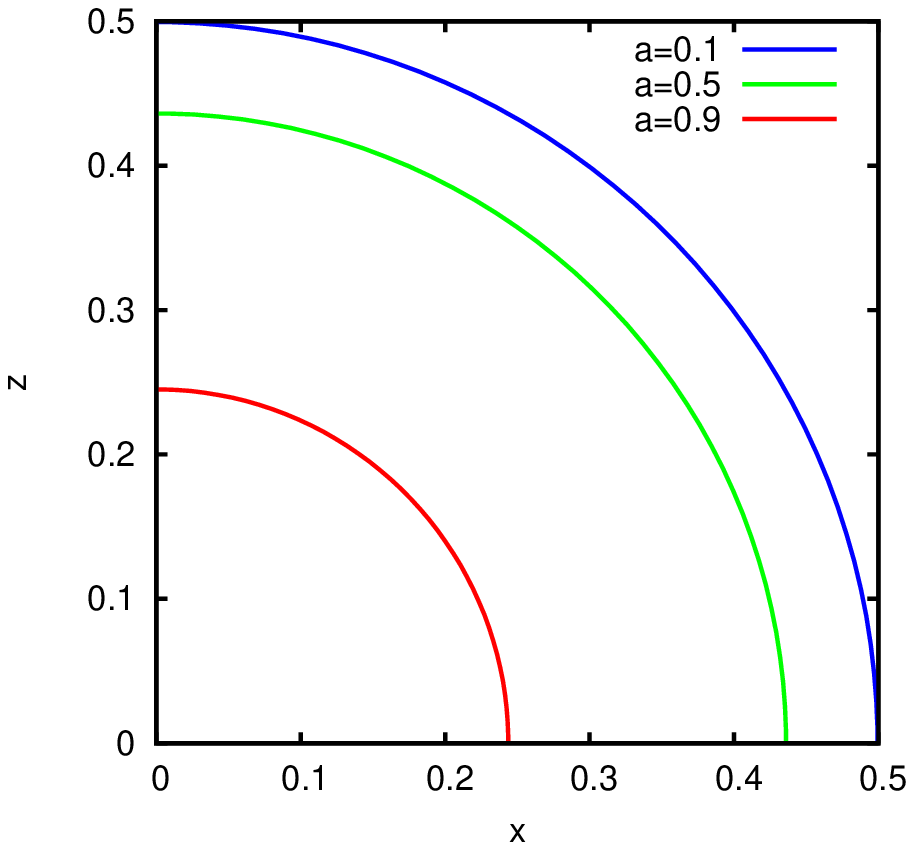,width=6.cm}\\
 (a)& (b)
\end{tabular}
 \vspace*{6pt}
 \caption{(a) AH surface of the Schwarzschild BH computed by the AH finder
 with SOR method.
 (b) The AH radius dependence of Kerr BH in isotropic coordinates by the spin parameter.
 \label{fig:AHF1}}
\end{figure}

In Fig.~\ref{fig:AHF1}~(a), we show the surface of AH of the
Schwarzschild BH in isotropic coordinates with the code ``sor\_AHF\_SBH\_ISO.f90''
and show the three dimensional shape of the AH in 1/8 spaces of
computational grid.
Fig.~\ref{fig:AHF1}~(b) shows the difference of the shape on x-z two
dimensional plane among different spin parameters with the code ``sor\_AHF\_KBH\_ISO.f90''.
The AH radius shrinks as the spin of BH increses.

\subsection{Kerr Black Hole and Single Puncture Black Hole}
As the last example, let us compare Kerr BH to single puncture BH
with a spin as initial data for numerical relativity.
A Kerr BH in quasi-isotropic coordinates can be used
as the initial data discussed in Sec.~\ref{sec:kerrBH}.
A single puncture BH is obtained by solving
the Hamiltonian constraint~(\ref{eq:ham_const_punc})
without any momenta $P^i=0$ and with a spin $S^z$ in the Bowen-York extrinsic
curvature~(\ref{eq:ext_by}).\\

\begin{figure}[ht]
\begin{tabular}{cc}
\psfig{file=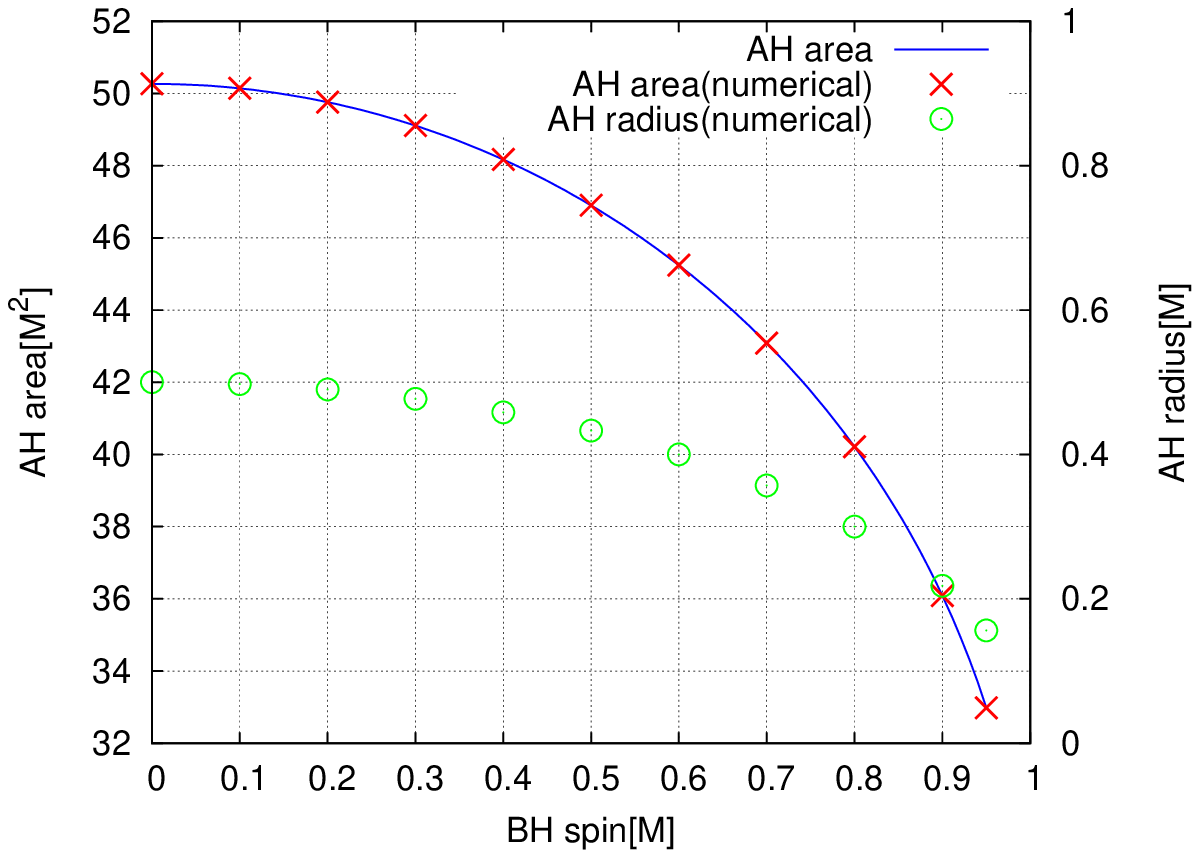,width=6.cm}&
\psfig{file=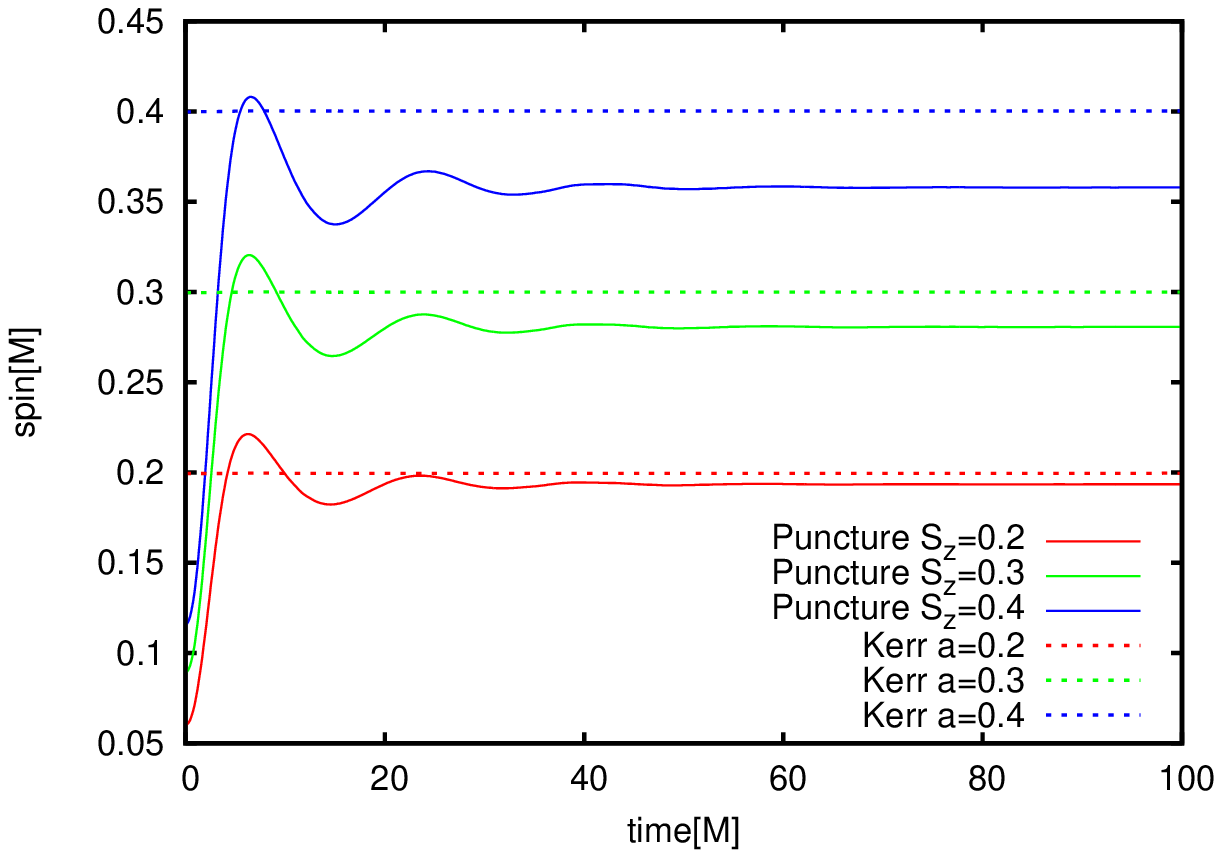,width=6.cm}\\
 (a)& (b)
\end{tabular}
 \vspace*{6pt}
 \caption{(a) The relation between the AH area and the spin of Kerr BH.
 The radius corresponding to the area in the quasi-isotropic coordinates
 is also shown.
 (b) The difference between puncture BH with a spin and Kerr BH.
 \label{fig:KBHvsPBH}}
\end{figure}

In order to check whether our AHF for this comparison works well, in Fig.~\ref{fig:KBHvsPBH}~(a)
we show the relation between AH area of the Kerr BH and
AH radius in isotropic coordinates 
as a function of spin parameter. The blue line denotes the analytical AH
area and red crosses denote numerical results by solving AH equation for
Kerr BH. The green circles show the coordinate radii where the AHs with
different are located.
Much larger computational resources should be required to obtain the solution with a high BH spin
because high resolution in the coordinate radius is required in this regime.\\

We perform numerical relativity simulations with the initial data of
single puncture BH and Kerr BH in Fig.~\ref{fig:KBHvsPBH}~(b).
The BSSN evolution equations which give stable dynamical evolution\cite{DavidLec,Shibata1995,Baumgarte1998}
are adopted in these simulations.
The color difference shows the difference among spins and
the type of lines denotes the difference between Kerr BH and single
puncture BH.
The spins of single puncture BHs settle down at late time,
which shows BHs relax to almost the stationary state
and one can compare results of Kerr BHs at late time.
The single puncture BH with the higher spin does not reach at the
spin which we expect.  This is because we assume the conformal flatness
for constructing puncture BH but Kerr BH should not be expressed by
the conformal flat metric.
However, it should be noted that the puncture BH can represent the small spin BH well
and it is actually powerful to construct the initial data for multi BHs system.

\section{Conclusions}
In these notes, we showed how to prepare the initial data for numerical
relativity and how to obtain the apparent horizon of BHs, which are reduced
to solving elliptic PDEs in general.
We presented several BH solutions as initial data for numerical
relativity and described several numerical methods to solve elliptic PDEs.
In particular, sample codes to solve Poisson equations
with linear and non-linear sources are available online to public users.
It is worth noting that these simple, ``classical'' methods are still powerful enough
to be of use for current problems.
In addition, we note that modern methods (e.g. Multi-Grid method)
can help to eventually upgrade these classical methods in terms of numerical costs and consuming-time.
Of course, one should carefully choose which method to use to solve elliptic PDEs, according to the problem
at hand. 

\section*{Acknowledgments}
The author would like to thank Vitor Cardoso for giving the opportunity
to lecture on this school, and to the Organizers and Editors of
the NR/HEP2: Spring School at Instituto Superior T\'ecnico in Lisbon.
The author would also thank an anonymous referee for a careful
reading of the manuscript and many useful suggestions.
The author is thankful to Ana Sousa who helps to improve English on this notes,
S\'ergio Almeida who maintains the cluster ``Baltasar-Sete-S\'ois''
and Takashi Hiramatsu who maintains the ``venus'' cluster.
Numerical computations in this work were carried out
on the cluster of ``Baltasar-Sete-S\'ois'' at Instituto Superior
T\'ecnico in Lisbon which is supported by the
DyBHo-256667 ERC Starting Grant and 
on the ``venus'' cluster at the Yukawa Institute Computer Facility in
Kyoto University.
This work was supported in part by Perimeter Institute for Theoretical Physics.
Research at Perimeter Institute is supported by the Government of Canada
through Industry Canada and by the Province of Ontario through the
Ministry of Economic Development \& Innovation.


\appendix

\section{LDU decomposition}
\label{app:LDUdecomp}
In this Appendix, we describe how to numerically solve only the Poisson equation.
However, we can also solve general elliptic PDEs in principle, namely,
in case except for the problem with diagonally dominant matrix.
A system of linear equations can be expressed by the matrix described in
Sec.~\ref{sec:matrix} as
\begin{eqnarray}
 A_{IJ}\psi_{J} &=& b_{I}.
\end{eqnarray}
Let us decompose a matrix $A_{IJ}$ into the lower and upper
triangular matrices defined as $L_{IJ}$ and $U_{IJ}$ respectively,
\begin{eqnarray}
 A_{IJ} &\equiv& L_{IK}D_{KK}U_{KJ}\nonumber\\
  &=& \left[
\begin{array}{ccccc}
 1 & 0 & 0 & \cdots & 0\\
 L_{21} & 1 & 0 & \cdots & 0\\
 L_{31} & L_{32} & 1 & \cdots & 0\\
 \vdots & \vdots & \vdots & \ddots & \vdots\\
 L_{N1} & L_{N2} & L_{N3} & \cdots & 1
\end{array}
    \right]
  \left[
\begin{array}{ccccc}
 D_{11} & 0 & 0 & \cdots & 0\\
 0 & D_{22} & 0 & \cdots & 0\\
 0 & 0 & D_{33} & \cdots & 0\\
 \vdots & \vdots & \vdots & \ddots & \vdots\\
 0 & 0 & 0 & \cdots & D_{NN}
\end{array}
    \right]
\left[
\begin{array}{ccccc}
 1 & U_{12} & U_{13} & \cdots & U_{1N}\\
 0 & 1 & U_{23} & \cdots & U_{2N}\\
 0 & 0 & 1 & \cdots & U_{3N}\\
 \vdots & \vdots & \vdots & \ddots & \vdots\\
 0 & 0 & 0 & \cdots & 1
\end{array}
    \right],\nonumber\\\label{eq:ludecompmatrix}
\end{eqnarray}
where $D_{KK}$ denotes the diagonal matrix.
Then, the solution vector can be solved $\psi_J$ step by step as
\begin{eqnarray}
b_I &=& A_{IJ}\psi_J = L_{IK}D_{KK}U_{KJ}\psi_J,\nonumber\\
&=& L_{IK}D_{KK}\xi_K,\label{eq:lowerxi}
\end{eqnarray}
where
\begin{eqnarray}
 \xi_K &\equiv& U_{KJ}\psi_J.\label{eq:upperpsi}
\end{eqnarray}
Thus, it is easy to obtain the solution as the following precedures.
First, we obtain an auxiliary vector $\xi_K$ as
\begin{eqnarray}
 \xi_{1} &=& b_{1}/D_{11},\nonumber\\
 \xi_{2} &=& \left(b_{2}-L_{21}D_{11}\xi_{1}\right)/D_{22},\nonumber\\
 \xi_{3} &=& \left(b_{3}-L_{31}D_{11}\xi_{1} -L_{32}D_{22}\xi_{2}\right)/D_{33},\nonumber\\
 \vdots & & \vdots\nonumber\\
 \xi_{N} &=& \left(b_{N} -\sum_{I=1}^{N-1}L_{NI}D_{II}\xi_{I}\right)/D_{NN},
\end{eqnarray}
solving Eq.~(\ref{eq:lowerxi}) from $\xi_{1}$ to $\xi_{N}$,
\begin{eqnarray}
 D_{11} \xi_{1} &=& b_{1},\nonumber\\
 L_{21}D_{11} \xi_{1} +D_{22}\xi_{2} &=& b_{2},\nonumber\\
 L_{31}D_{11} \xi_{1} +L_{32}D_{22}\xi_{2} +D_{33}\xi_{3} &=& b_{3},\nonumber\\
 \vdots & & \vdots\nonumber\\
 \sum_{I=1}^{N-1} L_{NI}D_{II}\xi_{I} +D_{NN}\xi_{N}&=& b_{N}.
\end{eqnarray}
Therefore, the solution vector $\psi_J$ is written by
\begin{eqnarray}
 \psi_{N} &=& \xi_{N},\nonumber\\  
 \psi_{N-1} &=& \xi_{N-1}-U_{N-1 N}\psi_{N},\nonumber\\
 \vdots & & \vdots\nonumber\\
 \psi_{1} &=& \xi_{1} -\sum_{I=N}^{2}U_{1I}\psi_{I},
\end{eqnarray}
similarly solving Eq.~(\ref{eq:upperpsi}) from $\psi_{N}$ to $\psi_{1}$,
\begin{eqnarray}
 \psi_{N} &=& \xi_{N},\nonumber\\
 U_{N-1 N} \psi_{N} +\psi_{N-1} &=& \xi_{N-1},\nonumber\\
 \vdots & & \vdots\nonumber\\
 \sum_{I=N}^{2} U_{1I}\psi_{I} +\psi_{1} &=& \xi_{1}.
\end{eqnarray}
As the last of this section, we note how we compute 
the lower and upper matrices from our matrix $A_{IJ}$, which is the time-consuming part.
The matrix $A_{IJ}$ is written with the diagonal, lower and upper
triangular matrices by
\begin{eqnarray}
 \begin{array}{ccrcclr}
  A_{IJ} &=& D_{IJ} &+&\sum_{K<I}L_{IK}D_{KK}U_{KJ}&,& diagonal\ (I=J),\\
  A_{IJ} &=& D_{II}U_{IJ} &+&\sum_{K<I}L_{IK}D_{KK}U_{KJ}&,& upper\ (I<J),\\ 
  A_{IJ} &=& L_{IJ}D_{JJ} &+&\sum_{K<J}L_{IK}D_{KK}U_{KJ}&,& lower\ (I>J).
 \end{array}
\end{eqnarray}
Thus, the components of marices are obtained in turn by
\begin{eqnarray}
  D_{II} &=& A_{II} - \sum_{J<I}L_{IJ}D_{JJ}U_{JI},\\
  U_{IJ} &=& \frac{1}{D_{II}} \left( A_{IJ} -\sum_{K<I}L_{IK}D_{KK}U_{KJ}\right),\\
  L_{IJ} &=& \frac{1}{D_{JJ}} \left( A_{IJ} -\sum_{K<J}L_{IK}D_{KK}U_{KJ}\right),
\end{eqnarray}
Although LDU decomposition allows us to numerically solve 
general elliptic PDEs,  the large numerical costs will be required
in many cases.

\section{Multi-Grid method}
\label{sec:MultiGrid}
Multi-Grid method is proposed by R.~Fedorenko and N.~Bakhvalov
and developed by A.~Brandt\cite{Fedorenko1962,Fedorenko1964,Bakhvalov1966,Brandt1977}.
The SOR method as mentioned in Sec.~\ref{sec:method_sor}
has the advantage of reducing the high frequency components
of residual between the exact solution and numerical solution,
because the values near the grid point to be updated are used
for next trial guess during the iteration.
On the other hand, it would take much time to reduce the low frequency modes of
redisual with this iteration method.
When we consider different resolution grids,
however, the low frequency modes on the finer grid
can be the high frequency modes on the coarser grid.
The low frequency modes of residual on the finer grid
can efficiently be reduced on the coarser grid.
The Multi-Grid method is based on the concept of
reducing different frequency modes of residual
with different resolution grids.
In fact, it was implemented by
some groups\cite{Choptuik1986,Hawley2003az,Brown2005}.

\subsection{Multi-Grid structure}
Suppose we have different resolution grids and the level of
different grids is labeled by $k$, which the larger $k$ denotes the finer grid.
One can solve the Poisson equation on the level $k$ by any iterative methods
described in Sec.~\ref{sec:numerical_methods}
and obtain the numerical solution,
\begin{eqnarray}
 \bigtriangleup^{(k)}\phi^{(k)} &=& S^{(k)},
\end{eqnarray}
where $\phi^{(k)}$ is the numerical solution on the level $k$.
We define the residual on the level $k$
between $\phi^{(k)}$ and the exact solution by
\begin{eqnarray}
 r^{(k)} &=& 
  S^{(k)} -\bigtriangleup^{(k)}\phi^{(k)}.
\end{eqnarray}

\subsubsection{Lagrange interpolation}
In general, the communication of the quantities such as the residual
with different grid levels is needed.
Now, we just use the Lagrange interpolation to communicate
with each other level defined by
\begin{eqnarray}
 F(x) &=& \sum_{j=0}^{N}F(x_j)L_j(x),\\
 L_j(x) &=& \prod_{i\neq j}^{N}\frac{x-x_i}{x_j-x_i},
\end{eqnarray}
where $F, x_j, x$ and $N$ denote
the quantity to be interpolated,
the coordinate on the level,
the location to be interpolated,
and the number of grid points to be used by the interpolation,
respectively.

\subsubsection{Restriction operator}
After we obtain the solution on the finer grid $k$,
we transfer the information of the solution from the finer grid $k$
to the coarser grid $k-1$.
Now we use the second-order discretization scheme and
choose the third-order Lagrange interpolation.
We define the modified source term on the coarser level $k-1$
with the information of the solution on the finer grid $k$ by
\begin{eqnarray}
 \phi^{(k-1)}_c &=& \mR^{k-1}_{k}\phi^{(k)},\\
 r^{(k-1)} &=& \mR^{k-1}_{k} r^{(k)},\\
 S^{(k-1)} &\equiv& \bigtriangleup^{(k-1)}\phi^{(k-1)}_c +r^{(k-1)}\nonumber\\
 &=& \bigtriangleup^{(k-1)} \left(\mR^{k-1}_{k}\phi^{(k)}\right)
  +\mR^{k-1}_{k}\left(S^{(k)} -\bigtriangleup^{(k)}\phi^{(k)}\right),\\
 \dif\phi^{(k-1)} &=& \phi^{(k-1)} -\phi^{(k-1)}_c,
\end{eqnarray}
where $\mR^{k-1}_{k}$ denotes the restriction operator
to the coarser grid $k-1$
and $\phi^{(k-1)}_c$ denotes the smoothing solution by the restriction operator.
Roughly speaking,
the modified source term $S^{(k-1)}$ consists of
that on the level $k$ with smoothing operation
and the correction by the difference of Laplacian operator
between two levels.
Then, we obtain the numerical solution $\phi^{(k-1)}$ on the level $k-1$
to solve the Poisson equation with the modified source term.

\subsubsection{Prolongation operator}
The solution with the modified source term on the coarser level $k-1$
is to be brought back to the finer level $k$.
Now the communication is also done by third-order Lagrange interpolation.
\begin{eqnarray}
 \phi^{(k)}_c &=& \mP^{k}_{k-1}\phi^{(k-1)},\\
 \dif\phi^{(k)}_c &=& \mP^{k}_{k-1}\dif\phi^{(k-1)} =
  \mP^{k}_{k-1}\left[\phi^{(k-1)}-\mR^{k-1}_{k}\phi^{(k)}\right],\\
\phi^{(k)}_m &\equiv& \phi^{(k)} +\dif\phi^{(k)}_c
  = \phi^{(k)} +  \mP^{k}_{k-1}\left[\phi^{(k-1)}-\mR^{k-1}_{k}\phi^{(k)}\right],\label{eq:set_trial_phi}\\
 \dif\phi^{(k)} &\equiv& \phi^{(k)}_m -\phi^{(k)}_c,\quad
  \phi^{(k)}=\phi^{(k)}_m,
\end{eqnarray}
where $\mP^{k}_{k-1}$ denotes the prolongation operator
and $\phi^{(k)}_m$ denotes the solution on the level $k$
modified by the coarser grid $k-1$.
The modification is done 
by Eq.~\eqref{eq:set_trial_phi}.

\subsubsection{Cycle of the Multi-Grid method}
There are some ways of deciding the order of the level to compute.
Fig.~(\ref{fig:cycle}) shows the difference of such order
between the methods of V-cycle and W-cycle as examples.
Now we choose V-cycle because it is easier to implement to the code.
We use the restriction operator
before computing on the coarser level 
and the prolongation operator 
before computing on the finer level.
This cycle is repeated until we obtain
the expected error of the Poisson equation.

\begin{center}
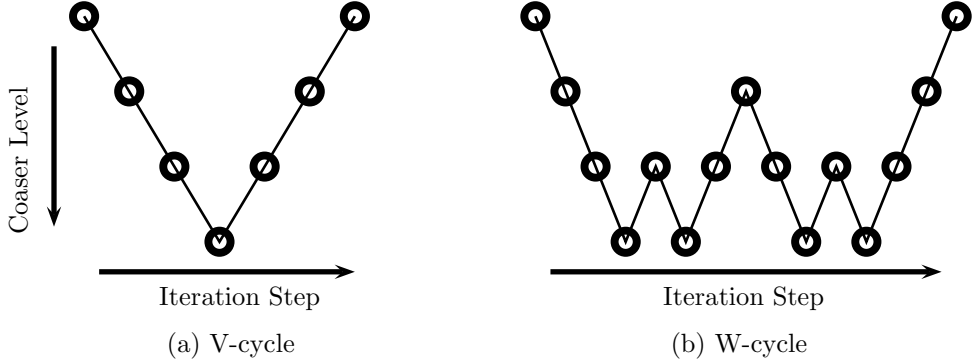
\begin{figure}[h]
\begin{minipage}[t]{0.8\textwidth}
\begin{pspicture}(0,0)(14,4)
 \pscircle[linewidth=3pt](1.,4.){0.2}
 \pscircle[linewidth=3pt](1.6,3.){0.2}
 \pscircle[linewidth=3pt](2.2,2.){0.2}
 \pscircle[linewidth=3pt](2.8,1.){0.2}
 \pscircle[linewidth=3pt](3.4,2.){0.2}
 \pscircle[linewidth=3pt](4.0,3.){0.2}
 \pscircle[linewidth=3pt](4.6,4.){0.2}
 \psline[linewidth=1pt](1.,4.)(1.6,3.)(2.2,2.)(2.8,1.)(3.4,2.)(4.,3.)(4.6,4.)
 \psline[linewidth=2pt]{->}(0.6,3.6)(0.6,1.2)
 \rput[lt]{90}(0.,1.1){Coaser Level}
 \psline[linewidth=2pt]{->}(1.2,0.6)(4.6,0.6)
 \rput[lt](2.,0.4){Iteration Step}
 \rput[lt](2.1,-0.2){(a) V-cycle}

 \pscircle[linewidth=3pt](7.,4.){0.2}
 \pscircle[linewidth=3pt](7.4,3.){0.2}
 \pscircle[linewidth=3pt](7.8,2.){0.2}
 \pscircle[linewidth=3pt](8.2,1.){0.2}
 \pscircle[linewidth=3pt](8.6,2.){0.2}
 \pscircle[linewidth=3pt](9.,1.){0.2}
 \pscircle[linewidth=3pt](9.4,2.){0.2}
 \pscircle[linewidth=3pt](9.8,3.){0.2}
 \pscircle[linewidth=3pt](10.2,2.){0.2}
 \pscircle[linewidth=3pt](10.6,1.){0.2}
 \pscircle[linewidth=3pt](11.,2.){0.2}
 \pscircle[linewidth=3pt](11.4,1.){0.2}
 \pscircle[linewidth=3pt](11.8,2.){0.2}
 \pscircle[linewidth=3pt](12.2,3.){0.2}
 \pscircle[linewidth=3pt](12.6,4.){0.2}
 \psline[linewidth=1pt](7.,4.)(7.4,3.)(7.8,2.)(8.2,1.)(8.6,2.)(9.,1.)(9.4,2.)(9.8,3.)
(10.2,2.)(10.6,1.)(11.,2.)(11.4,1.)(11.8,2.)(12.2,3.)(12.6,4.)
 \psline[linewidth=2pt]{->}(7.2,0.6)(12.4,0.6)
 \rput[lt](8.7,0.4){Iteration Step}
 \rput[lt](8.8,-0.2){(b) W-cycle}
\end{pspicture}
\vspace{0.2cm}
\end{minipage}
 \caption{Schematic picture of the Cycle. These are cases in which we have
 $4$ grid levels.}
 \label{fig:cycle}
\end{figure}
\end{center}

\subsection{Code test}
\begin{figure}[ht]
\begin{tabular}{cc}
\psfig{file=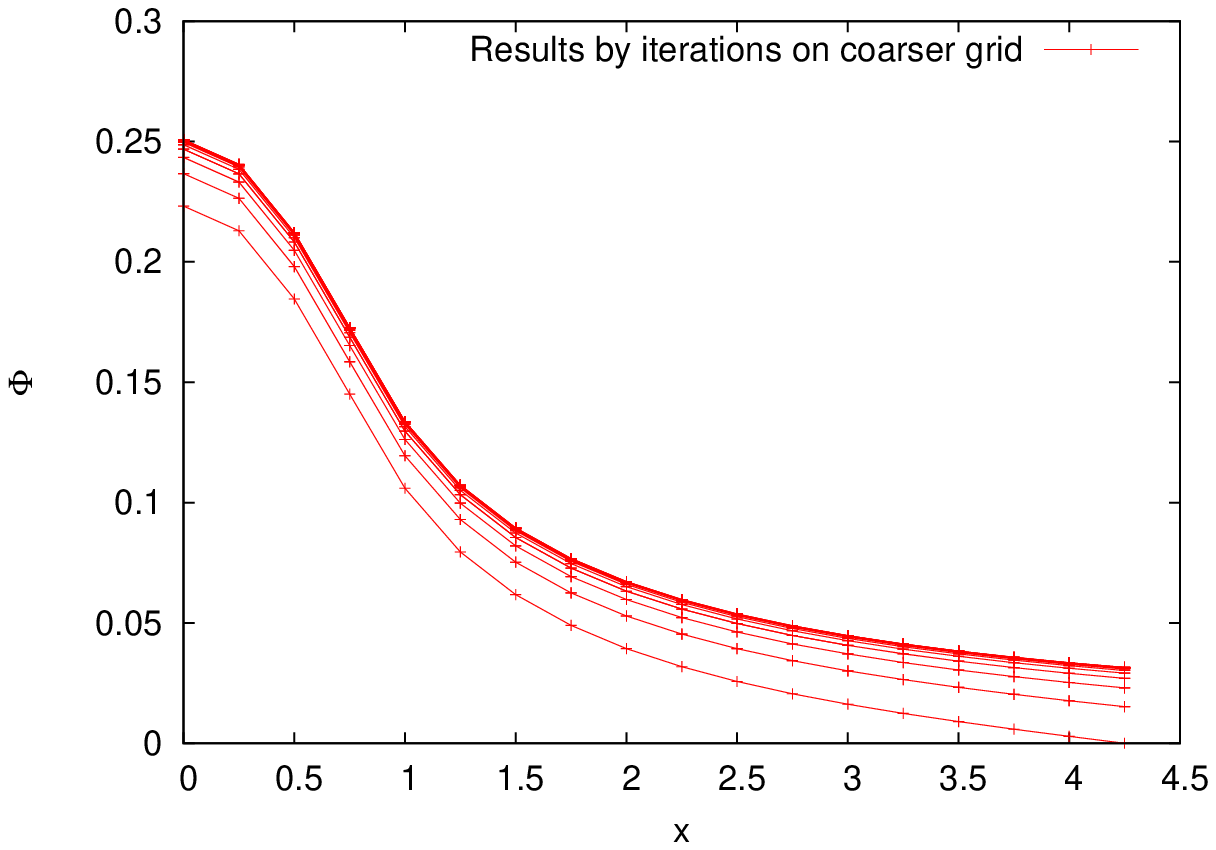,width=6.cm}&
\psfig{file=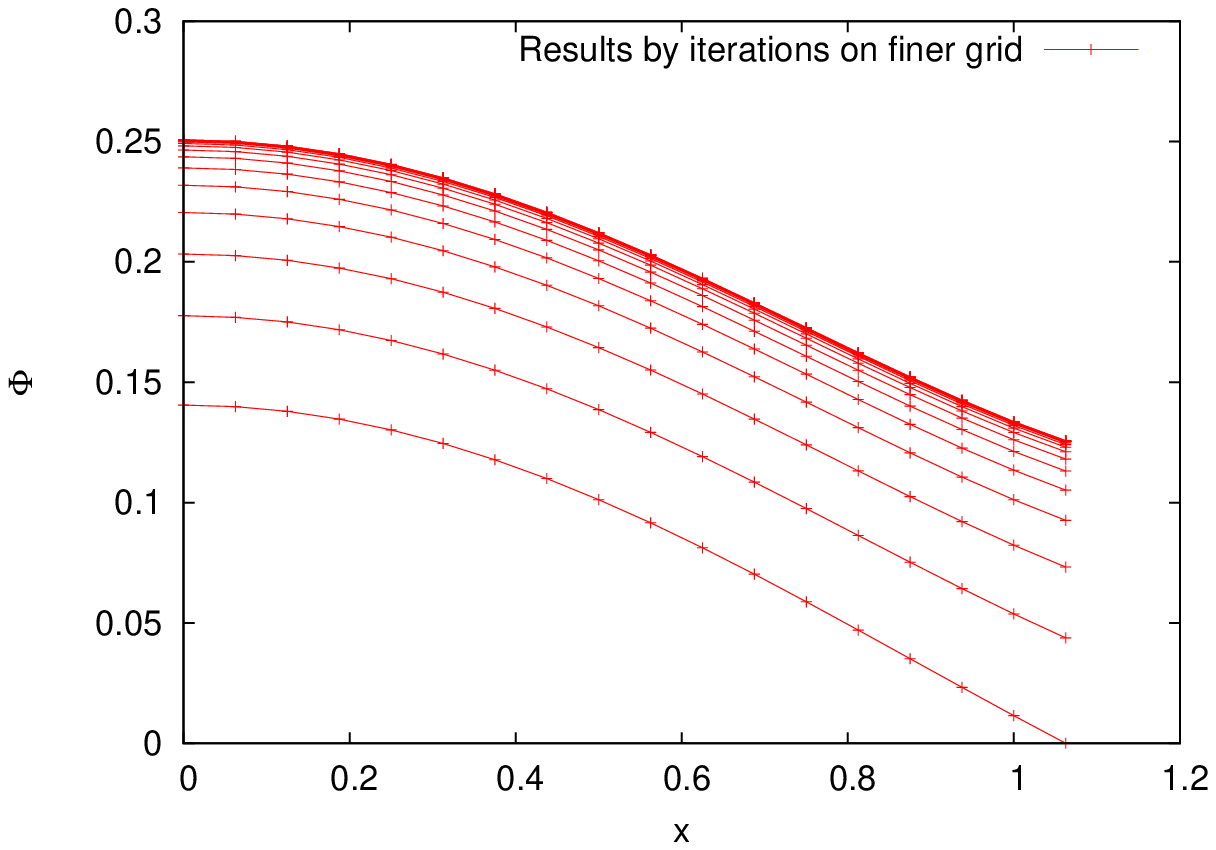,width=6.cm}\\
 (a)& (b)
\end{tabular}
 \vspace*{6pt}
 \caption{ The solution converges to the analytical solution discussed
 in Sec.~\ref{sec:test_nonlinear}. 
 (a) the finest grid level($k=3$).
 (b) the coarser grid level($k=1$).
 \label{fig:multigrid_test}}
\end{figure}

Let us consider the same test problem
as Sec.~\ref{sec:test_nonlinear}.
In the 3D problem, we impose the boundary conditions
at large distance by
\begin{eqnarray}
 0&=& \frac{\dif}{\dif r}(r\Phi)
 = \Phi +r\frac{\dif\Phi}{\dif r}
  =\Phi +x\frac{\del\Phi}{\del x} +y\frac{\del\Phi}{\del y}
  +z\frac{\del\Phi}{\del z}.
\end{eqnarray}
We note that the boundary of the finer grid is given by
the interpolation.  Fig.~\ref{fig:multigrid_test}
shows the results on the x-axis by solving the Poisson equation
with the source~\eqref{eq:source_grav} by Multi-Grid method.
The solution including the boundary converges to
the analytical solution by iterations.

\section{List of Sample codes}
\label{app:samplecode}
We have some sample codes for the lecture on NR/HEP2: Spring School at
Instituto Superior T\'ecnico in Lisbon and they are available online.
In this section, we show the simplest code to solve an elliptic PDE
and the sample code which is parallelized with OpenMP.  One can see what
is the parallel computing in Ref.~\cite{SergioLec}.
Here is the list of sample codes which are available in http://blackholes.ist.utl.pt/nrhep2/?page=material,
\begin{enumerate}
 \item jacobi\_test1.f90\\
       This is the code for solving the problem described in
       Sec.~\ref{sec:test_linear} with Jacobi method(See Sec.~\ref{sec:method_jacobi}).
 \item gs\_test1.f90\\
       This is the code for solving the problem described in
       Sec.~\ref{sec:test_linear} with Gauss-Seidel method(See Sec.~\ref{sec:method_gs}).
 \item sor\_test1.f90\\
       This is the code for solving the problem described in
       Sec.~\ref{sec:test_linear} with SOR method(See Sec.~\ref{sec:method_sor}).
 \item jacobi\_test2.f90\\
       This is the code for solving the problem described in
       Sec.~\ref{sec:test_linear} with Jacobi method(See Sec.~\ref{sec:method_jacobi}).
 \item sor\_AHF\_SBH\_ISO.f90\\
       This is the code for solving the AH of Schwarzschild BH with
       SOR method(See Sec.~\ref{sec:method_sor}).
 \item sor\_AHF\_KBH\_ISO.f90\\
       This is the code for solving the AH of Kerr BH in isotropic
       coordinates described in
       Sec.~\ref{sec:test_KBH} with SOR method(See Sec.~\ref{sec:method_sor}).
 \item sor\_AHF\_KBH\_BL.f90\\
       This is the code for solving the AH of Kerr BH in Boyer-Lindquist
       coordinates described in
       Sec.~\ref{sec:test_KBH} with SOR method(See Sec.~\ref{sec:method_sor}).
 \item jacobi\_openMP.f90\\
       This is the code for solving the problem described in
       Sec.~\ref{sec:test_linear} with Jacobi method(See Sec.~\ref{sec:method_jacobi})
       using many processors with OpenMP.
 \item jacobi\_test1.C\\
       This is the code written in C++ for solving the problem described in
       Sec.~\ref{sec:test_linear} with Jacobi method(See Sec.~\ref{sec:method_jacobi}).
 \item sor\_test1.C\\
       This is the code written in C++ for solving the problem described in
       Sec.~\ref{sec:test_linear} with SOR method(See Sec.~\ref{sec:method_sor}).
 \item jacobi\_openMP.C\\
       This is the code written in C++ for solving the problem described in
       Sec.~\ref{sec:test_linear} with Jacobi method(See Sec.~\ref{sec:method_jacobi})
       using many processors with OpenMP.
\end{enumerate}

\newpage
\subsection{jacobi\_test1.f90}
\footnotesize
\begin{verbatim}
     1	!@@@@@@@@@@@@@@@@@@@@@@@@@@@@@@@@@@@@@@@@@@@@@@@@@@@@@@@@@@@@@@
     2	!                                                              
     3	!     Jacobi method for TEST PROBLEM 1                         
     4	!                                                              
     5	!--------------------------------------------------------------
     6	!                                                              
     7	!     Sample Code for Lecture in NR/HEP2: Spring School         
     8	!                                                              
     9	!                                     Coded by Hirotada Okawa  
    10	!                                                              
    11	!==============================================================
    12	!   How to compile and use this program in terminal(bash)
    13	!==============================================================
    14	!   $ gfortran -O2 -ffast-math -o j_test1 jacobi_test1.f90
    15	!   $ ./j_test1
    16	!--------------------------------------------------------------
    17	!@@@@@@@@@@@@@@@@@@@@@@@@@@@@@@@@@@@@@@@@@@@@@@@@@@@@@@@@@@@@@@
    18	
    19	module inc_coord
    20	  implicit none
    21	
    22	!--------------------------------------------------------------
    23	!     Grid points
    24	!--------------------------------------------------------------
    25	  integer,parameter :: jli=1
    26	  integer,parameter :: jui=100
    27	  integer,parameter :: jlb=jli-1
    28	  integer,parameter :: jub=jui+1
    29	
    30	!--------------------------------------------------------------
    31	!     Maximum/Minimum Coordinates (physical)
    32	!--------------------------------------------------------------
    33	  real(8),parameter :: xlower=0.
    34	  real(8),parameter :: xupper=1.
    35	  real(8),parameter :: dx=(xupper-xlower)/dble(jub-jli)
    36	  real(8),parameter :: dxi=1.d0/dx
    37	
    38	!--------------------------------------------------------------
    39	!     Array for Coordinates (physical)
    40	!--------------------------------------------------------------
    41	  real(8),dimension(jlb:jub) :: x
    42	
    43	!--------------------------------------------------------------
    44	!     Variables to solve
    45	!--------------------------------------------------------------
    46	  real(8),dimension(jlb:jub) :: h, hprev
    47	
    48	!--------------------------------------------------------------
    49	!     Source term for Poisson equation
    50	!--------------------------------------------------------------
    51	  real(8),dimension(jlb:jub) :: src
    52	
    53	end module inc_coord
    54	
    55	program main
    56	  use inc_coord
    57	  implicit none
    58	
    59	!--------------------------------------------------------------
    60	!     Definition of parameters
    61	!--------------------------------------------------------------
    62	  integer,parameter :: stepmax=1d8      ! Loop step maximum
    63	  real(8),parameter :: errormax=1.d-10  ! Error to exit loop
    64	  real(8),parameter :: fpar=1.d0        ! for next guess
    65	
    66	!--------------------------------------------------------------
    67	!     Definition of temporary variables to use
    68	!--------------------------------------------------------------
    69	  integer :: j, step
    70	  real(8) :: xx
    71	  real(8) :: errortmp,vtmp
    72	
    73	!--------------------------------------------------------------
    74	!     Output File
    75	!--------------------------------------------------------------
    76	  open(200,file='h_j.dat')
    77	
    78	!--------------------------------------------------------------
    79	!     Initialization
    80	!--------------------------------------------------------------
    81	  do j=jlb,jub
    82	     x(j)  = xlower +(dble(j)-0.5d0)*dx    ! Coordinates
    83	     h(j)  = 1.d0                          ! variable to solve
    84	     hprev(j) = h(j)                       ! previous variable
    85	  end do
    86	
    87	!**************************************************************
    88	!     Main Loop
    89	!**************************************************************
    90	  do step=0,stepmax
    91	
    92	!--------------------------------------------------------------
    93	!     Preserve data of previous step
    94	!--------------------------------------------------------------
    95	     do j=jlb,jub
    96	        hprev(j) = h(j)
    97	     end do
    98	
    99	!--------------------------------------------------------------
   100	!     Jacobi Method
   101	!--------------------------------------------------------------
   102	     do j=jli,jui
   103	
   104	!==============================================================
   105	!     Definition of Source term
   106	!==============================================================
   107	        xx   = x(j)
   108	        src(j) = xx**2*12.
   109	
   110	        h(j) = 0.5d0*( hprev(j+1) +hprev(j-1) -dx**2*src(j) )
   111	     end do
   112	
   113	!==============================================================
   114	!     Impose Boundary Condition
   115	!==============================================================
   116	     h(jub)=x(jub)**4      ! Dirichlet Boundary Condition
   117	     h(jlb)=h(jli)         ! Neumann Boundary Condition
   118	
   119	!--------------------------------------------------------------
   120	!     Check if values converge
   121	!--------------------------------------------------------------
   122	     errortmp=0.d0
   123	     vtmp=0.d0
   124	     do j=jli,jui
   125	        errortmp = errortmp +(h(j)-hprev(j))**2*dx**2
   126	        vtmp = vtmp + dx**2
   127	     end do
   128	     errortmp = dsqrt(errortmp/vtmp)
   129	     if( (errortmp.le.errormax) .and. (step.gt.1) ) exit
   130	
   131	!--------------------------------------------------------------
   132	!     Next Guess
   133	!--------------------------------------------------------------
   134	     do j=jlb,jub
   135	        h(j) = fpar*h(j) +(1.d0-fpar)*hprev(j)
   136	     end do
   137	
   138	     write(*,*) "Step=",step,"Error=",errortmp
   139	  end do
   140	
   141	!--------------------------------------------------------------
   142	!     Print Data
   143	!--------------------------------------------------------------
   144	  do j=jlb,jub
   145	     write(200,'(4e16.8e2)') x(j),h(j),hprev(j),src(j)
   146	  end do
   147	
   148	!--------------------------------------------------------------
   149	!     End of Program
   150	!--------------------------------------------------------------
   151	  write(*,*) "End of Run",errortmp
   152	  close(200)
   153	
   154	end program main
\end{verbatim}

\newpage
\subsection{jacobi\_openMP.f90}
\footnotesize
\begin{verbatim}
     1	!@@@@@@@@@@@@@@@@@@@@@@@@@@@@@@@@@@@@@@@@@@@@@@@@@@@@@@@@@@@@@@@@@@@@
     2	!                                                                    
     3	!     Jacobi method for TEST PROBLEM 1                               
     4	!                             parallelized with OpenMP               
     5	!                                                                    
     6	!--------------------------------------------------------------------
     7	!                                                                    
     8	!     Sample Code for Lecture in NR/HEP2: Spring School               
     9	!                                                                    
    10	!                                     Coded by Hirotada Okawa        
    11	!                                                                    
    12	!====================================================================
    13	!   How to compile and use this program in terminal(bash)
    14	!====================================================================
    15	!   $ gfortran -O2 -ffast-math -fopenmp -o j_omp jacobi_openMP.f90
    16	!   $ export OMP_NUM_THREADS=2
    17	!   $ ./j_omp
    18	!--------------------------------------------------------------------
    19	!   OMP_NUM_THREADS : Change the number of cores you want to use.
    20	!--------------------------------------------------------------------
    21	!@@@@@@@@@@@@@@@@@@@@@@@@@@@@@@@@@@@@@@@@@@@@@@@@@@@@@@@@@@@@@@@@@@@@
    22	
    23	module inc_coord
    24	  implicit none
    25	
    26	!--------------------------------------------------------------
    27	!     Grid points
    28	!--------------------------------------------------------------
    29	  integer,parameter :: jli=1
    30	  integer,parameter :: jui=100
    31	  integer,parameter :: jlb=jli-1
    32	  integer,parameter :: jub=jui+1
    33	
    34	!--------------------------------------------------------------
    35	!     Maximum/Minimum Coordinates (physical)
    36	!--------------------------------------------------------------
    37	  real(8),parameter :: xlower=0.
    38	  real(8),parameter :: xupper=1.
    39	  real(8),parameter :: dx=(xupper-xlower)/dble(jub-jli)
    40	  real(8),parameter :: dxi=1.d0/dx
    41	
    42	!--------------------------------------------------------------
    43	!     Array for Coordinates (physical)
    44	!--------------------------------------------------------------
    45	  real(8),dimension(jlb:jub) :: x
    46	
    47	!--------------------------------------------------------------
    48	!     Variables to solve
    49	!--------------------------------------------------------------
    50	  real(8),dimension(jlb:jub) :: h, hprev
    51	
    52	!--------------------------------------------------------------
    53	!     Source term for Poisson equation
    54	!--------------------------------------------------------------
    55	  real(8),dimension(jlb:jub) :: src
    56	
    57	end module inc_coord
    58	
    59	program main
    60	  use inc_coord
    61	  implicit none
    62	
    63	!--------------------------------------------------------------
    64	!     Definition of parameters
    65	!--------------------------------------------------------------
    66	  integer,parameter :: stepmax=1d8      ! Loop step maximum
    67	  real(8),parameter :: errormax=1.d-10  ! Error to exit loop
    68	  real(8),parameter :: fpar=1.d0        ! for next guess
    69	
    70	!--------------------------------------------------------------
    71	!     Definition of temporary variables to use
    72	!--------------------------------------------------------------
    73	  integer :: j, step
    74	  real(8) :: xx
    75	  real(8) :: errortmp,vtmp
    76	
    77	!--------------------------------------------------------------
    78	!     Output File
    79	!--------------------------------------------------------------
    80	  open(200,file='h_o.dat')
    81	
    82	
    83	!--------------------------------------------------------------
    84	!     OpenMP threads folk
    85	!--------------------------------------------------------------
    86	!$OMP PARALLEL DEFAULT(SHARED) PRIVATE(j,xx,src)
    87	
    88	!--------------------------------------------------------------
    89	!     Initialization
    90	!--------------------------------------------------------------
    91	!$OMP DO
    92	  do j=jlb,jub
    93	     x(j)  = xlower +(dble(j)-0.5d0)*dx    ! Coordinates
    94	     h(j)  = 1.d0                          ! variable to solve
    95	     hprev(j) = h(j)                       ! previous variable
    96	  end do
    97	!$OMP END DO
    98	
    99	!**************************************************************
   100	!     Main Loop
   101	!**************************************************************
   102	  do step=0,stepmax
   103	
   104	!--------------------------------------------------------------
   105	!     Preserve data of previous step
   106	!--------------------------------------------------------------
   107	!$OMP DO
   108	     do j=jlb,jub
   109	        hprev(j) = h(j)
   110	     end do
   111	!$OMP END DO
   112	
   113	!--------------------------------------------------------------
   114	!     Jacobi Method
   115	!--------------------------------------------------------------
   116	!$OMP DO
   117	     do j=jli,jui
   118	
   119	!==============================================================
   120	!     Definition of Source term
   121	!==============================================================
   122	        xx   = x(j)
   123	        src(j) = xx**2*12.
   124	
   125	        h(j) = 0.5d0*( hprev(j+1) +hprev(j-1) -dx**2*src(j) )
   126	     end do
   127	!$OMP END DO
   128	
   129	!==============================================================
   130	!     Impose Boundary Condition
   131	!==============================================================
   132	!$OMP SINGLE
   133	     h(jub)=x(jub)**4      ! Dirichlet Boundary Condition
   134	     h(jlb)=h(jli)         ! Neumann Boundary Condition
   135	!$OMP END SINGLE
   136	
   137	!--------------------------------------------------------------
   138	!     Check if values converge
   139	!--------------------------------------------------------------
   140	     errortmp=0.d0
   141	     vtmp=0.d0
   142	!$OMP BARRIER
   143	!$OMP DO REDUCTION(+:vtmp,errortmp)
   144	     do j=jli,jui
   145	        errortmp = errortmp +(h(j)-hprev(j))*(h(j)-hprev(j))*dx**2
   146	        vtmp = vtmp + dx**2
   147	     end do
   148	!$OMP END DO
   149	
   150	!$OMP SINGLE
   151	     errortmp = dsqrt(errortmp/vtmp)
   152	!$OMP END SINGLE
   153	     if( (errortmp.le.errormax) .and. (step.gt.1) ) exit
   154	
   155	!--------------------------------------------------------------
   156	!     Next Guess
   157	!--------------------------------------------------------------
   158	!$OMP DO
   159	     do j=jlb,jub
   160	        h(j) = fpar*h(j) +(1.d0-fpar)*hprev(j)
   161	     end do
   162	!$OMP END DO
   163	
   164	!$OMP SINGLE
   165	     write(*,*) "Step=",step,"Error=",errortmp
   166	!$OMP END SINGLE
   167	  end do
   168	
   169	!--------------------------------------------------------------
   170	!     OpenMP threads join
   171	!--------------------------------------------------------------
   172	!$OMP END PARALLEL
   173	
   174	!--------------------------------------------------------------
   175	!     Print Data
   176	!--------------------------------------------------------------
   177	  do j=jlb,jub
   178	     write(200,'(4e16.8e2)') x(j),h(j),hprev(j),src(j)
   179	  end do
   180	
   181	!--------------------------------------------------------------
   182	!     End of Program
   183	!--------------------------------------------------------------
   184	  write(*,*) "End of Run",errortmp
   185	  close(200)
   186	
   187	end program main
\end{verbatim}

\newpage

\bibliographystyle{ws-ijmpa}
\bibliography{bib_okawa}

\end{document}